\documentclass[prd,twocolumn,superscriptaddress,floatfix,nofootinbib]{revtex4-2}
\pdfoutput=1 
\usepackage[T1]{fontenc}
\usepackage{amsmath}
\usepackage{amssymb}
\usepackage{amsthm}
\usepackage{amsfonts}
\usepackage{graphicx}
\usepackage{enumitem}
\usepackage{bm}
\usepackage{rotating}
\usepackage{hyperref}
\usepackage{pbox}
\usepackage[dvipsnames]{xcolor}
\usepackage{array}
\usepackage{physics}
\usepackage{dsfont}
\usepackage{mathtools}
\usepackage{leftidx}
\usepackage{lmodern}
\usepackage[normalem]{ulem}
\usepackage{braket}
\usepackage{tensor}
\usepackage{mathrsfs}
\usepackage{float}
\usepackage{dsfont}
\usepackage[margin=2cm]{geometry}
\usepackage[title]{appendix}
\usepackage{tcolorbox}

\usepackage{tikz}
\usetikzlibrary{arrows.meta,decorations.pathmorphing,math}

\newcommand{\tcb}[1]{\leavevmode{\color{Blue}{#1}}}

\renewcommand{\bar}{\overline}
\renewcommand{\tilde}{\widetilde}

\newcommand{\ii}{\mathsf{i}}

\newcommand{\sx}{\mathsf{x}}
\newcommand{\sy}{\mathsf{y}}

\newcommand{\rr}[1]{\left(#1\right)}
\newcommand{\bx}{{\bm{x}}}

\newcommand{\bk}{{\bm{k}}}

\newcommand{\supp}{\text{supp}}

\newcommand{\rao}{\hat{\rho}_{\textsc{a}}^{0}}
\newcommand{\rbo}{\hat{\rho}_{\textsc{b}}^{0}}

\newcommand{\rof}{\hat{\rho}_{\phi}^{0}}

\newcommand{\roo}{\hat{\rho}^{0}}

\newtheorem*{lemma}{Lemma}

\newcommand{\R}{\mathbb{R}}
\newcommand{\C}{\mathbb{C}}
\newcommand{\M}{\mathcal{M}}

\newcommand{\A}{\mathcal{A}}
\newcommand{\W}{\mathcal{W}}
\newcommand{\CS}{C^\infty_0(\M)}
\newcommand{\Sol}{\mathsf{Sol}}

\begin{document}

\title{Quantum teleportation with relativistic communication from first principles}

\author{Erickson Tjoa}
\email{e2tjoa@uwaterloo.ca}
\affiliation{Department of Physics and Astronomy, University of Waterloo, Waterloo, Ontario, N2L 3G1, Canada}
\affiliation{Institute for Quantum Computing, University of Waterloo, Waterloo, Ontario, N2L 3G1, Canada}

\begin{abstract}
    In this work we provide a genuine relativistic quantum teleportation protocol whose classical communication component makes use of relativistic causal propagation of a quantum field. Consequently, the quantum teleportation is fully relativistic by construction. Our scheme is based on Unruh-DeWitt qubit detector model, where the quantum state being teleported is associated to an actual qubit rather than a field mode  considered by Alsing and Milburn [\href{https://journals.aps.org/prl/abstract/10.1103/PhysRevLett.91.180404}{PRL \textbf{91}, 180404 (2003)}]. We show that the existing works in (relativistic) quantum information, including good definitions of one-shot and asymptotic channel capacities, as well as algebraic formulation of quantum field theory, already provide us with all the necessary ingredients to construct  fundamentally relativistic teleportation protocol in relatively straightforward manner.
    
\end{abstract}

\maketitle

\section{Introduction}

Quantum teleportation \cite{Bennett1993teleport} is undoubtedly one of the simplest yet most remarkable discoveries in the field of quantum information, which has also been experimentally realized \cite{bouwmeester1997experimental,Popescu1998teleportexperiment}. The protocol shows, in particular, that one can in effect transmit one unit of quantum information (qubit) by communicating two classical bits and consuming a unit of shared entanglement  (``ebit''), namely a Bell state. The original proposal by Bennett \textit{et. al.} \cite{Bennett1993teleport} assumes perfect protocol at every step: the two parties only need to have the following: a state to be teleported, a shared Bell pair, a classical communication channel, and a conditional unitary operation by Bob to ``telecorrect'' his state after obtaining Alice's measurement outcome. There is no need to consider any dynamics or time evolution in the protocol (this is separate from any actual experimental implementations, which do depend on how they are set up). In particular, the protocol is manifestly non-relativistic.

In the last two decades, some attempts have been made to incorporate some aspects of relativity, such as the effect of noise induced by non-trivial motion in spacetime. This was first considered by Alsing and Milburn \cite{Alsing2003teleport}, where one of the parties undergoes uniformly accelerated motion. The key component in such investigations is the concept of entanglement associated to observers in different (noninertial) frames (also see, e.g., \cite{Fuentes2005entanglement,Alsing2012review,Hu2012review}). However, this protocol has conceptual issues: in particular, the use of field modes as a ``qubit'' is unphysical due to its highly delocalized nature and the cavity setup presents some problems regarding how to account for the teleportation fidelity, as pointed out in \cite{Schutzhold2005critique}. A better proposal was shortly given by Landulfo and Matsas \cite{Landulfo2009suddendeath}, where actual qubit detector model based on the construction by Unruh and Wald \cite{Unruh1984detector} was used, now more commonly known as the \textit{Unruh-DeWitt} (UDW) detector model \cite{Unruh1979evaporation,DeWitt1979}. This proposal recognizes the fact that the state to be teleported should be physically localized in spacetime and realizable in principle, for which a qubit traveling in spacetime is suitable for.

Our work is motivated by the fact that the nice proposal in \cite{Landulfo2009suddendeath} is somewhat incomplete, since it still assumes that the local operations and classical communication (LOCC) part of the protocol can be performed somehow and respects relativity ``by default''. Depending on the purpose, we can argue that if anything, it is the CC part that should be relativistic since Alice and Bob can ``protect'' their shared Bell pair before the protocol begins. This is the case, for instance, if we would like to consider teleportation protocol in curved spacetime where Alice and Bob place themselves somewhere in spacetime with all their local laboratories protected, and then perform the teleportation protocol using \textit{quantum field} as a means of classical communication. Viewing this in another way, the construction in \cite{Landulfo2009suddendeath,Hu2012review} is really about what happens to teleportation when all qubits are immersed in a quantum field where Unruh effect can manifest: it is more about the impact of noise coming from interacting with a relativistic ``environment'' (the field) when one detector undergoes relativistic motion through the field, rather than about  relativistic version of the teleportation protocol. Indeed the original teleportation protocol does not consider environmental-induced decoherence \textit{as part of the protocol}\footnote{This is distinct from the fact that any \textit{practical implementation} of the protocol needs to fight against environmental noise.}.

In this work we close this gap in relativistic quantum information (RQI) by proposing a genuine relativistic quantum teleportation protocol. Our protocol is based on the minimal requirement that at the fundamental level, all the \textit{relevant} components for the protocol must respect relativistic principles, i.e.,
\begin{enumerate}[leftmargin=*,label=(\alph*)]
    \item Alice and Bob's quantum systems must (at least) be \textit{somewhere} in spacetime and has spatiotemporal degrees of freedom that are consistent with relativity. The bare minimum for this to work is for the quantum system to have classical motional degree of freedom that moves along timelike trajectories, which the UDW detector model can furnish.
    
    \item The classical communication should be implemented by a relativistic field (classical or quantum) by construction. This ensures that relativistic causality is respected from first principles, rather than an assumption as is done in standard textbooks on quantum information (under the name ``no superluminal signalling''). 
    
    \item The quantum state to be teleported must be physically accessible. For example, the construction in \cite{Alsing2003teleport} is not physical because each mode is completely delocalized over all spacetimes. Furthermore, one cannot perform projective measurements on quantum fields \cite{josepolo2022measurement}, so any protocol that tries to do Bell measurements on field modes directly is out of the window.
\end{enumerate}
While our protocol is unlikely to be useful for practical purposes in quantum computation (since there are likely more efficient and clever ways to do this from experimental perspectives), our goal is to show that from theoretical and fundamental standpoints, there exists relativistically covariant quantum teleportation protocol that fully respects causality \textit{by construction}. The practical implementation can thus be regarded as a ``coarse-grained''  protocol where relativity is no longer necessary for the user to be aware of\footnote{For example, in \cite{bouwmeester1997experimental} or \cite{Popescu1998teleportexperiment} all we need to accept is that the ``photons'' used in the setup are excitations of the electromagnetic field, whose theory is consistent with relativity. In other words, components like  optical fibres respect relativity as a black box, since the experiment can be performed without any real relativistic calculations.}.

The relativistic teleportation protocol we propose here is also based on the UDW detector model, similar to the approach used in \cite{Landulfo2009suddendeath}, but we consider the case when the spacetime is an arbitrary globally hyperbolic Lorentzian manifold $(\M,g_{ab})$ \tcb{where $g_{ab}$ is the spacetime metric}. Furthermore, since the Bell measurement and the ``telecorrecting'' unitary are modeled as being performed instantaneously (very quickly), we will use a delta-coupling approach where the coupling with the field is very rapid, effectively at a single instant in time. The delta-coupling model has been used in the study of relativistic quantum communication, specifically regarding the ability of a quantum channel to transmit one classical or quantum bit (see, e.g., \cite{Cliche2010channel,jonsson2018transmitting,Simidzija2020transmit,tjoa2022channel}). It has also been used to study other protocols such as entanglement harvesting or lack thereof \cite{Gallock-Yoshimura2021mutualinfo,Henderson2020temporal,Simidzija2018nogo} and sabotaging correlations \cite{sahu2021sabotaging}.

As shown in \cite{tjoa2022channel}, the delta-coupling approach readily fits into the algebraic approach to quantum field theory, which provides us with much flexibility and generality. As such, our relativistic protocol constitutes an algebraic reformulation of the quantum teleportation that respects relativity. Such reformulation provides greater unity and economy, since many calculations simplify in the algebraic framework (essentially due to the Weyl relations), and furthermore the results obtained will be valid in arbitrary curved spacetimes \textit{and} any of the Hilbert space representations of the quantum field.

We stress that in essence what needs to be changed to accommodate a fully relativistic quantum teleportation protocol is the classical communication part. If we simply assume that the LOCC component is perfect and magical (i.e., it succeeds \textit{somehow}), then quantum teleportation protocol (or really, any of the standard quantum protocols in quantum information theory) is completely independent of whether relativity theory holds. In fact, any background theory which has finite signal propagation speed is sufficient: one simply has to acknowledge that before Alice's signal reaches Bob, Bob cannot know the measurement outcome and the output will be maximally mixed. If we consider non-relativistic theory of spacetime \textit{a la} Newton, the speed of light is effectively $c\to\infty$, so the only difference is that Bob can do the measurement immediately after Alice's measurement and outcomes sent (instantaneously). In an even more extreme case, one can consider purely quantum-information-theoretic causal structure of spacetime without assuming any relativistic spacetime at all (see, e.g., \cite{vilasini2022embedding,Paunkovic2020causalorders} and refs. therein): all one needs is a concept of causal partial ordering and signalling partial ordering.

This paper is organized as follows. In Section~\ref{sec: teleportation-protocol} we review the kinematics of teleportation, one-shot and asymptotic channel capacities, and then provide a simple relativistic extension. In Section~\ref{sec: extension} we comment on some possible extensions and variants. \tcb{In Appendix~\ref{sec: AQFT} we provide the  bare minimum of the algebraic framework for quantization of scalar field theory for interested readers}. We adopt mostly-plus signature for the metric and use natural units $c=\hbar=1$.

\section{Teleportation protocol}
\label{sec: teleportation-protocol}

In this section we describe a fully relativistic quantum teleportation protocol. We first review the ``kinematical'' aspect of the teleportation protocol, which only depends on the LOCC acting on Hilbert spaces; we then propose a replacement of the CC component with a fully relativistic channel.

\subsection{The kinematics of teleportation}
In the standard teleportation protocol, Alice and Bob are assumed to share three copies of two-dimensional Hilbert spaces $\mathcal{H} = \mathcal{H}_{A_1}\otimes\mathcal{H}_{A}\otimes\mathcal{H}_{B}$. The first copy is for Alice's qubit whose state is to be teleported; the second and third copy is for Alice and Bob to share a Bell state (as a resource). The teleportation protocol uses this shared entanglement resource and 2 bits of classical information to effectively give a perfect quantum channel that can transmit 1 bit of quantum information. Note that the protocol says nothing about how to implement the classical communication  part.

The basic idea of the teleportation protocol is as follows. First, consider the joint initial state to be
\begin{align}
    \ket{\psi} &= \ket{\zeta_{00}}\ket{\Phi^+} \,,\quad \ket{\zeta_{00}} = \alpha\ket{0}+\beta\ket{1}\,,
\end{align}
where $\alpha,\beta \in \C\,$, $|\alpha|^2+|\beta|^2 = 1$. The notation $\ket{\zeta_{00}}$ will become clear in what follows. As usual, the trick is to rewrite the first three sectors in a way that the entanglement in sector $AB$ is swapped into sector $A_1A$:
\begin{align}
    \hspace{-0.25cm}\ket{\zeta_{00}}\ket{\Phi^+} 
    &= \frac{1}{2}\ket{\Phi^+}\ket{\zeta_{00}}+ \frac{1}{2}\ket{\Phi^-}\ket{\zeta_{01}}
    \notag\\
    &+\frac{1}{2}\ket{\Psi^+}\ket{\zeta_{10}}+\frac{1}{2}\ket{\Psi^-}\ket{\zeta_{11}}\,,\\
    \ket{\zeta_{00}} &\coloneqq \alpha\ket{0}+ \beta\ket{1}\,,\quad \ket{\zeta_{01}}  \coloneqq \alpha\ket{0} - \beta\ket{1}\notag\\
    \ket{\zeta_{10}} &\coloneqq \beta\ket{0} + \alpha\ket{1}\,,\quad \ket{\zeta_{11}}  \coloneqq \beta\ket{0} - \alpha\ket{1}\,.
\end{align}
Here the four Bell states are
\begin{align}
    \ket{\Phi^\pm} &= \frac{\ket{00}\pm\ket{11}}{\sqrt{2}}\,,\quad
    \ket{\Psi^\pm} = 
    \frac{\ket{01}\pm \ket{10}}{\sqrt{2}}\,.
\end{align}
Alice now wants to perform Bell measurement associated to sector $A_1A$. The four possible outcomes can be associated to four messages $\{00,01,10,11\}$ respectively (hence we labelled $\zeta_{ij}$ this way). The standard teleportation protocol then requires that Bob applies a conditional unitary on his share of Bell state according to the message:
\begin{equation}
    \begin{aligned}
    00: \ket{\zeta_{00}}&\mapsto \openone\ket{\zeta_{00}} \,,\\
    01: \ket{\zeta_{01}}&\mapsto \hat\sigma^z\ket{\zeta_{01}} \,,\\
    10: \ket{\zeta_{10}}&\mapsto \hat\sigma^x\ket{\zeta_{10}} \,,\\
    11: \ket{\zeta_{11}}&\mapsto \hat\sigma^y\ket{\zeta_{11}} \,.
\end{aligned}
\label{eq: control-unitary}
\end{equation}
In all cases, the unitary will produce the state $\ket{\zeta_{00}}$ that Alice originally possesses. Note that Alice does \textit{not} know the quantum state (i.e., the value of $\alpha,\beta$), since otherwise we would not need the teleportation protocol and simply inform Bob to prepare the state separately --- teleportation is one of the so-called `oblivious' protocols \cite{Wilde2013textbook}. 

Quantum teleportation is considered to be one of the three simplest quantum information protocols that involve communication between two parties --- the other two being superdense coding and entanglement distribution \cite{Wilde2013textbook}.  However, in all the three protocols every component except the communication part does not actually rely on relativistic principles. Within their own laboratories, Alice and Bob only need to have a set of qubits and shared (entanglement) resources to perform the task, as well as local operators within their own laboratory (but they are allowed to perform Bell measurements \tcb{within their own labs}). In fact more is true: the protocols are effectively only \textit{kinematical} in the sense that they are only about mapping between Hilbert spaces. This is the basis for the distinction between the notion of relativistic causality and \textit{information-theoretic causality} \cite{Paunkovic2020causalorders,vilasini2022embedding}. The latter only relies on some order relation between events associated to some random variables between them. 

The important interplay between the two notions of causality is the fact that information-theoretic causality has to be compatible with relativistic causality if the background spacetime structure exists, but the former is strictly more general since it cannot be fully described classically. This is because information-theoretic causality  has to provide a quantum-mechanical notion of causal structure that captures essential features of non-classicality in Einstein-Podolsky-Rosen and Bell-type experiments. The distinction between the two notions of causality has led to a wealth of research, most notably for developing quantum frameworks to model causal relations and causation (see, e.g.,\cite{Leifer2013bayesian,Leofer2006dynamics,Wood2015causal,Pienaar2015causal,ried2015quantum,Costa2016causal,Fritz2012beyond}) and more recently for understanding indefinite causal orders\footnote{This is indefinite causal ordering in the sense of information-theoretic causality. It is currently debatable whether this can be made into a concrete realization of indefinite relativistic causal ordering: see \cite{vilasini2022embedding} for most recent work trying to disentangle these two concepts, notably Section 7 and 8.} and causal inequalities (see, e.g., \cite{Hardy2007towards,zych2019bell,oreshkov2012quantum,Chiribella2013computation,Arajo2015nonseparable,Oreshkov2016separable,Chiribella2021comm,Vedral2020fridge}).

In view of the above considerations, there is a sense in which quantum information protocols usually described in standard textbooks \cite{Wilde2013textbook,nielsen2000quantum} have no real dynamical content\footnote{We are not talking about any experimental implementation, which of course has dynamics in it. However, even there relativity is basically absent from the basic description (since it is not directly relevant for understanding the outcomes).} because they only pertain to information-theoretic notion of causality, not relativistic causality. Furthermore, one does not have to even talk about time evolution --- the qubits involved in the protocols do not even need to obey quantum mechanical evolution given by the Schr\"odinger equation (but obey the rest of the postulates such as Born's rule, measurement updates, linearity and superposition). Both special relativity and general relativity are theories prescribing how physical systems behave in spacetime. Consequently, a quantum information protocol that manifestly respects relativity must include spacetime as part of its description\footnote{We are excluding quantum gravity proposals which may have classical relativistic limits, it is a story better told another time.} --- in other words, we need to be able to describe where the physical systems are located, their causal relations, and how they impact the protocol's ability to perform the task. Clearly, the only way relativity enters the picture is through the communication channel, and it is this step that needs to be modified. Our task is to describe how to do this in the subsequent sections.

\subsection{Classical communication via quantum channel}

At the core of the teleportation protocol, Alice needs to communicate of her (Bell) measurement outcome to Bob. This step is usually done in standard quantum information essentially ``by declaration'': Alice can always (somehow) send two bits of classical information to Bob for adjusting the type of unitary to apply to Bob's share of Bell state. Certainly, the current experimental implementation shows that this can indeed be done, and clearly it must respect relativistic principle at the fundamental level. However, this makes the role of relativity almost invisible or irrelevant from the framework (i.e., e-mails and phone calls are ``relativistic black boxes''), so from fundamental standpoint this is not ideal. Here we outline two types of communication protocols: {one-shot} and {asymptotic} settings (see \cite[Chp. 7]{Wilde2013textbook} and refs. therein for more details).

The starting point is to suppose that Alice has a set of messages $M$ that she wishes to communicate to Bob. Each message $m$ can be associated to a bit string of at least size $\log_2|M|$: for example, in teleportation case we have four distinct messages and the bare minimum is to have them written as $M=\{00,01,10,11\}$, where each message $m$ is a bit string of length $2$. Alice has an encoder, i.e., a list of input variables  $\{x_i: i\in M\}$ called a codebook of size $|M|$. In this case, the index corresponds to one of the four messages Alice wishes to send. Bob has a corresponding decoder that implements a POVM $\{E_{j}\}$ which contains at least $|M|$ POVM elements; the decoder outputs another index $j$ such that decoding error is said to have occurred $j\neq i$. Bob's success probability depends on the ``quality'' of the channel, which we make more precise below. We assume that both Alice and Bob know the sort of channel they will be using, so they can agree on the codebook and choose the appropriate encoding/decoding schemes for the task.

Let us first consider the communication channel in one-shot setting. For teleportation purposes the message will be deterministic, thus it is convenient to describe Alice's initial state to be given by a classically correlated state\footnote{In general Alice does not have to send a deterministic message, so she can consider instead $\rho^{A_1A_2} = \sum_{m}p_m\ketbra{m}{m}_{A_1}\otimes \ketbra{m}{m}_{A_2}$, which reduces to the deterministic case by picking $p_{m'} = \delta_{m',m}$ \cite{Wilde2013textbook}.}
\begin{align}
    \rho_{m}^{A_1A_2} &= \ketbra{m}{m}\otimes \ketbra{m}{m}\,,\quad m\in M\,.
\end{align}
where $\rho_{m}^{A_1A_2}$ is a density operator acting on $\mathcal{H}_{A_1}\otimes\mathcal{H}_{A_2}$ and each $\mathcal{H}_{A_j}\cong \C^2\otimes \C^2$. The extra copy in $A_1$ is meant to keep track of the message, and for our purposes we do not distinguish $x_i$ from $i$ because the messages are already naturally in the binary representation. For example, if Alice wishes to send $01$ then she prepares a classically correlated state $\ketbra{01}{01}\otimes\ketbra{01}{01}$. Alice encodes her message into qubit $A_2$ using a classical-quantum (CQ) channel $\mathcal{E}:\mathscr{D}(\mathcal{H}_{A_2})\to \mathscr{D}(\mathcal{H}_{A_2})$ given by \cite{Wilde2013textbook} 
\begin{align}
    \mathcal{E}(\ketbra{m}{m'}) &= \delta_{m,m'}\rho_{m}^{A_2}\,,
\end{align}
where the message $m$ is associated to some state $\rho_m^{A_2}$. The state after encoding is then given by
\begin{align}
    \tilde{\rho}^{A_1A_2}_m &\coloneqq   \ketbra{m}{m}\otimes  \mathcal{E}(\ketbra{m}{m})=\ketbra{m}{m}\otimes  \rho_{m}^{A_2}\,.
\end{align}
Alice sends the state $\rho_{m}^{A_2}$ through a communication channel $\Phi:\mathscr{D}(\mathcal{H}_{A_2})\to\mathscr{D}(\mathcal{H}_{B_1})$. The total state becomes
\begin{align}
    \sigma^{A_2B_1}_m &\coloneqq  \ketbra{m}{m} \otimes  \Phi(\rho_{m}^{A_2})\,.
    \label{eq: final-channel-after}
\end{align}
Bob then performs a decoding measurement using some POVM $\{E_{m}\}$, which can be described as a quantum-classical  channel $\mathcal{D}:\mathscr{D}(\mathcal{H}_{B_1})\to \mathscr{D}(\mathcal{H}_{B_1})$ such that the total state now reads
\begin{align}
    \tilde\sigma^{A_2B_1}_{m,m'} &\coloneqq (\openone\otimes\mathcal{D})(\sigma^{A_2B_1}_m) \notag\\
    &= \ketbra{m}{m}\otimes \sum_{m'\in M} \tr(E_{m'}\Phi(\rho_{m}^{A_2}))\ketbra{m'}{m'}\,,
\end{align}
Bob declares that Alice's message was $m$ if Bob's measurement corresponds to the POVM element $E_{m}$. In practice this means that the conditional probability
\begin{align}
    \Pr(m'=m|m) & = \tr(E_{m}\Phi(\rho_{m}^{A_2})) 
\end{align}
should be large enough to be useful (i.e., error probability is low). Once Bob decodes the message, Bob simply performs the correct unitary on qubit $B$ (Bob's share of the Bell state). Notice that in this protocol, the success probability of teleporting the state is inevitably dependent on the success of the communication protocol, even if the unitary can be applied perfectly.


The rate at which classical information can be transmitted via (asymptotically) large number of independent uses of the channel $\Phi$ is given by the famous Holevo-Schumacher-Westmoreland (HSW) channel coding theorem \cite{holevo1998capacity,schumacher1997sending}. Generalizations without the assumption of each use being independent (``memoryless'') were given in \cite{hayashi2003general,kretschmann2005quantum}. However, these results are strictly speaking only true for the limit of infinitely many uses of the channel with error probability taken to zero.  In this sense, ``one-shot channel capacity'' is desirable because it only concerns the number of bits transmitted in a single use of the channel for a given error probability. Furthermore, such capacity will include the HSW theorem because large independent uses of the channel $\Phi$ is the same as a single product channel $\Phi^{\otimes n}$.

In \cite{Renner2012oneshot} the so-called $\epsilon$-one-shot channel capacity of an arbitrary quantum channel was defined, with tight bounds defined in terms of relative-entropy-type quantity. More precisely, consider the \textit{hypothesis testing} of distinguishing two possible states of a system $\rho_1,\rho_2$ using two POVM elements $P$ and $Q=\openone-P$. The probability of guessing correctly the input state $\rho_1$ is $\tr(Q\rho_1)$, while the wrong guess is $\tr(Q\rho_2)$. The \textit{hypothesis testing relative entropy}\footnote{This is related to other forms of ``smoothened'' relative Reny\'i entropy: see also \cite{Buscemi2010oneshot,Datta2013oneshotassist,Datta2013smooth,Anshu2019oneshot}.} is defined to be
\begin{align}
    D^\epsilon_H(\rho_1||\rho_2)&= -\log_2\hspace{-0.2cm}\inf_{\substack{0\leq Q\leq \openone\\
    \tr(Q\rho_1)\geq 1-\epsilon}}\hspace{-0.2cm}\tr(Q\rho_2)\,,
\end{align}
which can be efficiently computed via semidefinite programming. From the definition the error probability of this discrimination task is given by $\epsilon\in [0,1)$. The hypothesis testing relative entropy $D^\epsilon_H(\cdot\,||\,\cdot)$ obeys analogous properties of relative entropy such as monotonocity and positivity, and is related to the relative entropy via quantum Stein's lemma \cite{hiai1991proper}:
\begin{align}
    D(\rho||\sigma) = \lim_{n\to\infty}\frac{1}{n}D^\epsilon_H(\rho^{\otimes n}||\sigma^{\otimes n})\,.
\end{align}
Following \cite{Renner2012oneshot}, let us define $\pi^{AB}\in \mathscr{D}(\mathcal{H}_A\otimes\mathcal{H}_B)$ by
\begin{align}
    \pi^{AB}\coloneqq \sum_{m}p_M(m)\ketbra{m}{m}_A\otimes \rho_m^{B}\,,
\end{align}
where $p_M$ is the probability distribution associated to random variable $M$ for which Eq.~\eqref{eq: final-channel-after} is a special case (identifying $A_2=A$ and $B_1=B$). Define $\pi^A,\pi^B$ to be the corresponding marginal states of $\pi^{AB}$. The one-shot capacity $\mathfrak{C}$ is the largest number for which a code of length $2^\mathfrak{C}$ exists with error probability $\epsilon$ satisfying the inequality
\begin{align}
    \mathfrak{C}_{\text{min}} &\leq \mathfrak{C}\leq \mathfrak{C}_{\text{max}}
    \label{eq: one-shot-capacity}
    \,,
\end{align}
where
\begin{subequations}
\begin{align}
    \mathfrak{C}_{\min} &= \sup_{p_M} D^{\epsilon/2}_H(\pi^{AB}||\pi^A\otimes\pi^B) -\log_2\frac{1}{\epsilon}-4 \,,\\
    \mathfrak{C}_{\max} &= \sup_{p_M} D^\epsilon_H(\pi^{AB}||\pi^A\otimes\pi^B)\,.
\end{align}
\end{subequations}
In general $\mathfrak{C}_{\min}>0$ and hence $\mathfrak{C}>0$, i.e., the code $(2^{\mathfrak{C}},\epsilon)$ exists, and some approximations of $\mathfrak{C}$ is also available in terms of smoothened min- and max-entropies \cite{Renner2012oneshot}. In general, however, the bounds are unwieldy.

In the asymptotic scenarios, Alice and Bob are allowed to make $n$ independent uses of the communication channel where $n$ is large. This actually amounts to one-shot scenario but for the product channel $\Phi^{\otimes n}$ \cite{Wilde2013textbook}. Alice can still use a product state as the input across all independent uses of the channel, so that her new encoding channel is a map $\mathcal{E}':\mathscr{D}(\mathcal{H}_{\textsc{a}})\to \mathscr{D}(\mathcal{H}_{\textsc{a}}^{\otimes n})$ such that
\begin{align}
    \mathcal{E}'(\ketbra{m}{m'}) &= \delta_{m,m'}\rho_{m}^{A_2^{\otimes n}}\,.
\end{align}
Alice applies $\Phi^{\otimes n}:\mathscr{D}(\mathcal{H}_{\textsc{a}}^{\otimes n})\to \mathscr{D}(\mathcal{H}_{\textsc{b}}^{\otimes n})$ to $\mathcal{E}'(\ketbra{m}{m'})$, representing $n$-independent uses of the communication channel $\Phi$. Bob then uses an extended decoding channel $\mathcal{D}': \mathscr{D}(\mathcal{H}_{\textsc{b}}^{\otimes n})\to \mathscr{D}(\mathcal{H}_{\textsc{b}})$, giving the resulting state
\begin{align}
    \hspace{-0.25cm}\tilde{\rho}^\textsc{b}_m &\coloneqq  \sum_{m'\in M} \tr(\tilde{E}_{m'}(\Phi^{\otimes n}\circ\mathcal{E}')(\ketbra{m}{m}))\ketbra{m'}{m'}\,,
\end{align}
where $\{\tilde{E}_m\}$ is the extended POVM across multiple uses of the channel. The error probability is now
\begin{align}
    \tilde{\Pr}(m'=m|m) & = \tr(\tilde{E}_{m}(\Phi^{\otimes n}\circ\mathcal{E}')(\ketbra{m}{m}))\,.
\end{align}
Hence, the one-shot setting includes the asymptotic setting as a special case. The {rate} $R$ of the communication channel is defined to be the number of bits transmitted averaged over the number of independent uses of the channel:
\begin{align}
    R=\frac{1}{n}\log_2|M|\,.
\end{align}
If the classical communication has maximum error probability (of Bob's decoding measurement) $\epsilon\geq 0$, we say that we have a code $(n,R,\epsilon)$. We say that the rate $R$ is \textit{achievable} if for any $\delta,\epsilon>0$ there exists a code $(n,R-\delta,\epsilon)$ for sufficiently large $n$. 

The HSW theorem \cite{holevo1998capacity,schumacher1997sending} states that the channel capacity of $n$ independent uses of the channel is given by \cite{Wilde2013textbook,nielsen2000quantum}
\begin{align}
    C(\Phi) &= \lim_{n\to\infty}\frac{1}{n}\chi(\Phi^{\otimes n})\,,
    \label{eq: capacity-asymptotic}
\end{align}
where $\chi$ is the Holevo information, defined by the maximization over all ensembles of input states $\{p_m,\rho_m\}$ \cite{nielsen2000quantum}:
\begin{align}
    \chi(\Phi) &= \max_{\{p_m,\rho_m\}}S\Bigr(\Phi\bigr(\sum_m p_m\rho_m\bigr)\Bigr) - \sum_m p_m S(\Phi(\rho_m))\,.
\end{align}
If Alice encodes her message into product states across each $n$ independent use of $\Phi$, then the channel capacity actually reduces to just the Holevo information, also known as \textit{product state capacity} $C^{(1)}$ \cite{nielsen2000quantum}:
\begin{align}
    C^{(1)}(\Phi) = \chi(\Phi)\,.
\end{align}
However, if Alice allows for entangled input across $\Phi^{\otimes n}$, the asymptotic limit \eqref{eq: capacity-asymptotic} is needed and this is typically intractable except for some special classes of channels. Note that the HSW does imply that the product state channel capacity is always nonzero except for the replacement channel $\Phi(\rho) = \rho_0$ for some fixed $\rho_0$.

Fortunately, the relativistic quantum channel we will consider falls under the class of \textit{entanglement-breaking channels}, where the Holevo information of the product channel is additive, i.e. $\chi(\Phi^{\otimes n}) = n\chi(\Phi)$, so that the channel capacity is also equal to its Holevo quantity:
\begin{align}
    C(\Phi) = \chi(\Phi)\,.
\end{align}
The main takeaway of this subsection is that the channel we will consider in the next subsection will have in general nonzero capacity to transmit classical information, be it in the one-shot or asymptotic, and for the asymptotic case the result has been more or less established in the non-perturbative regime for both the delta-coupled and gapless detectors \cite{Landulfo2016magnus1,tjoa2022channel}.

\subsection{Relativistic classical communication}

Our main goal is to show that the classical communication component of the teleportation protocol can be described exclusively via a simple UDW-type interaction between a qubit and a quantum field, even in the case when the interaction is isotropic in spacetime. In other words, Alice uses a quantum channel $\Phi$ induced by the field to communicate two bits of information by ``symmetric wireless broadcasting''. Such model of classical communication has two advantages: (1) it is valid in arbitrary globally hyperbolic curved spacetimes, and (2) relativistic causality is manifest from first principle by construction. 

One conceptually straightforward way of accommodating relativistic classical communication is to imagine Alice having \textit{three} qubits and Bob having \textit{two} qubits, and they share a quantum field $\phi$ mediating the communication: the total Hilbert space is a tensor product of six distinct Hilbert spaces:
\begin{align}
    \mathcal{H} &= \underbrace{\mathcal{H}_{A_1}\otimes\mathcal{H}_{A}\otimes\mathcal{H}_{B} }_{\text{kinematical}}\otimes\,\underbrace{\mathcal{H}_{A_2}\otimes\mathcal{H}_{B_1}\otimes\mathcal{H}_\phi}_{\text{communication }}\,.
    \label{eq: six-hilbert-spaces}
\end{align}
The idea is that we assume Alice and Bob can initialize and protect their qubits in the ``kinematical sector'' $\mathcal{H}_{\textsc{kin}}\equiv \mathcal{H}_{A_1}\otimes\mathcal{H}_{A}\otimes\mathcal{H}_{B}$ perfectly (since they can simply perform state preparation at the very beginning). Then Alice and Bob communicate using two UDW detectors interacting with quantum field in the ``communication sector'' $\mathcal{H}_{\textsc{comm}}\equiv\mathcal{H}_{A_2}\otimes\mathcal{H}_{B_1}\otimes\mathcal{H}_\phi$. For our purposes, the field's Hilbert space is with respect to the GNS representation of the vacuum state (which is a quasifree state). 

Crucially, it is worth stressing that the above setup is different from the ``relativistic'' teleportation protocol considered in the literature \cite{Alsing2003teleport,Landulfo2009suddendeath} because there what is relativistic is the ``environment'' and the states of motion of Alice and Bob's qubits. The fact that the teleportation fidelity decreases in general is simply a manifestation that detector-field interaction decreases purity of the qubits; furthermore, states of motion of both detectors and local curvature of spacetimes generically provide additional sources of noise that impact the fidelity even further. Therefore, strictly speaking there is nothing relativistic about the protocol itself: one would get the same result if one replaces the field with a lattice of quantum harmonic oscillators and consider non-relativistic trajectories. In contrast, for genuine relativistic teleportation protocol, it is \textit{essential} that (at least) the CC part of the teleportation protocol is relativistic: local curvature and states of motion of the detector are secondary contributions that can be made consistent with the description of CC component. 

Our task is to construct a quantum channel $\Phi$ that is fundamentally relativistic and is able to transmit classical information from Alice to Bob. One way to do this is to use the sort of quantum channel considered in \cite{tjoa2022channel} (or \cite{Landulfo2016magnus1,Landulfo2021cost} if we use gapless detector), which we review below.

We suppose that Alice's qubit $A_2$ and Bob's qubit $B_1$ are associated to a UDW detector --- for convenience we will drop the subscript and call them $A$ and $B$. The UDW detector is a two-level system (``a qubit'') with free Hamiltonian given by
\begin{align}
    \mathfrak{h}_j &= \frac{\Omega_j}{2}(\hat\sigma^z_j+\openone)\,,\quad j=A,B
\end{align}
where $\hat{\sigma}^z_j$ is the usual Pauli-$Z$ operator for detector $j$, whose ground and excited states $\ket{g_j},\ket{e_j}$ have energy $0,\Omega_j$ respectively. Let $\tau_j$ be the proper time of detector $j$ whose centre of mass travels along the worldline $\sx_j(\tau_j)$. \textit{A priori} the proper times may not coincide (i.e., the sense that $\dd\tau_A/\dd\tau_B\neq 1$) due to relativistic redshift caused by relative motion or different gravitational potential.

The covariant generalization of the UDW model is first given by \cite{Tales2020GRQO}: the  interaction Hamiltonian four-form (in interaction picture) reads
\begin{align}
    h_{I,j}(\sx) &= \dd V\,f_j(\sx)\hat\mu_j(\tau_j(\sx))\otimes\hat\phi(\sx)\,,
    \label{eq: hamiltonian-four-form}
\end{align}
where $\dd V = \dd^4\sx \sqrt{-g}$ is the invariant volume element in $\M$, $f_j(\sx)\in\CS$ prescribes the interaction region between detector $j$ and the field. The monopole moment of the detector $j$, denoted $\hat\mu_j(\tau_j)$, is given by
\begin{align}
    \hat\mu_j(\tau_j) &= \hat \sigma^x_j(\tau_j) = \hat\sigma^+_j e^{\ii\Omega_j \tau_j} + \hat\sigma^-_je^{-\ii\Omega_j \tau_j}\,,
\end{align}
where $\hat\sigma^\pm$ are the $\mathfrak{su}(2)$ ladder operators with $\hat\sigma^+_j\ket{g_j}=\ket{e_j}$ and $\hat\sigma^-_j\ket{e_j}=\ket{g_j}$. 

The total unitary time-evolution for the detector-field system is given by the time-ordered exponential with respect to some time parameter $t$ (in interaction picture) \cite{Bruno2020time-ordering}:
\begin{align}
    U = \mathcal{T}_t\exp\left[-\ii \int_\M \dd V h_{I,A}(\sx)+h_{I,B}(
    \sx) \right]\,.
\end{align}
The issue of time ordering of such unitary has been investigated in \cite{Bruno2020time-ordering}.

The delta-coupling regime for the UDW model is the regime where the interaction timescale between the detector and the field is assumed to be much faster than all the relevant timescales of the problem.  This regime is particularly suited for the analysis of the teleportation problem. For our purposes, the way to handle this is to consider the delta-coupling detector in terms of Fermi normal coordinates\footnote{For much more comprehensive discussions on Fermi normal coordinates, its applications and limitations, see \cite{Tales2022FNC}.} (FNC), denoted by $(\tau,\bar{\bx})$ so that we have
\begin{align}
    f_j(\sx) = \lambda_j\eta_j\delta(\tau_j-\tau_{j,0})F_j(\bar{\bx})\label{eq: delta-smearing-analytic}\,,
\end{align}
Since $\supp f_\textsc{a}\cap\supp f_\textsc{b}=\emptyset$, there is no problem using local FNCs associated to each trajectory. From global hyperbolicity, there exists a global time function $t$ coming from the foliation $\M\cong \R\times \Sigma_t$ so we are able to make sense of time ordering between the two interaction regions\footnote{Note that this can be really difficult in general unless the trajectories have some symmetries, since there is a lot of arbitrariness in how to embed Cauchy slices in $\M$.}. If the trajectories are stationary so that the proper times are proportional to coordinate time $t$, then the time ordering is clear since we can align $t$ with the proper times of the detectors.

The unitary $U$ reduces to a product of two simple (ordered) unitaries $U=U_{\textsc{b}}U_{\textsc{a}}$, where
\begin{align}
    U_{j} &= \exp\left[-\ii \hat\mu_j(\tau_{j,0}) \otimes \hat Y_j \right]\,,\label{eq: unitary-delta}\\
    \hat Y_j &= \hat\phi(f_j) = \int_\M\dd V\,f_j(\sx)\hat\phi(\sx) \,,
    \label{eq: smeared-Y-free}
\end{align}
with $f_j(\sx)$ is given in Eq.~\eqref{eq: delta-smearing-analytic}. This unitary can be written as a sum of bounded operators
\begin{align}
    U_j = \openone \otimes \cos\hat Y_j-\ii\hat\mu_j(\tau_{j,0})\otimes\sin\hat Y_j\,.
\end{align}
Again here $\tau_{j,0}$ can be ordered using the global time coordinate since we will be mostly interested in stationary trajectories (for which teleportation is our focus), noting that some additional care required if the trajectory of each detector is non-trivial (e.g., accelerating detectors)\footnote{Again, the problem is that we can always have good time ordering between two detectors by choosing suitable Cauchy surfaces, but the one that works best may not coincide with the global time coordinate most naturally associated to the spacetime of interest.}. 

\tcb{In what follows, we will exploit the fact that the operators $\cos\hat\phi(f_j),\sin\hat\phi(f_j)$ are bounded operators in the algebra of observables $\A(\M)$ of the quantum field, thus the framework of algebraic quantum field theory (AQFT) is very convenient (see Appendix~\ref{sec: AQFT} for more details). That said,} the results of this work using delta-coupled detector is not strictly speaking at the level of rigour of pure mathematics since the spacetime smearing $f_j(\sx)$ for delta coupling in Eq.~\eqref{eq: delta-smearing-analytic} does not belong to $\CS$. Intuitively, delta coupling corresponds to very rapid interactions, i.e., an approximation of smooth compactly supported functions that are very localized in time. The use of AQFT in this work is mainly to ensure that the setting generalizes easily to arbitrary curved spacetimes and for any of the GNS representations, and such coupling seems natural for the teleportation protocol.

In what follows we drop the subscript in $A_2,B_1$ and write $A,B$ to remove clutter since we are working strictly only in the communication sector $\mathcal{H}_\textsc{comm}$. Let us consider the initial state of the detector-field system to be given by initially uncorrelated state
\begin{align}
    \roo &= \rao \otimes \rbo \otimes \rof\,.
\end{align}
We are interested in the quantum channel $\Phi:\mathscr{D}(\mathcal{H}_\textsc{a})\to \mathscr{D}(\mathcal{H}_\textsc{b})$, where $\mathscr{D}(\mathcal{H}_j)$ is the space of density matrices associated to Hilbert space $\mathcal{H}_j$ of detector $j$. The channel is naturally defined in the Stinespring-like representation
\begin{align}
    \Phi(\rao) = \tr_{\textsc{b},\phi}(U\roo U^\dagger)\,. 
\end{align}
We can calculate the channel explicitly in closed form: it is given by \cite{Tjoa2022fermi}
\begin{widetext}
\begin{align}
    &\hat{\rho}_{\textsc{ab}\phi} = U(\rao\otimes\rbo\otimes\rof)U^\dagger \notag\\
    &\!\!= \rbo\otimes\rr{\rao\otimes C_\textsc{b}C_\textsc{a}\rof C_\textsc{a}C_\textsc{b} +
    \hat\mu_\textsc{a}\rao\hat\mu_\textsc{a}\otimes C_\textsc{b}S_\textsc{a}\rof S_\textsc{a}C_\textsc{b} +
    \ii\rao\hat\mu_\textsc{a}\otimes C_\textsc{b}C_\textsc{a}\rof S_\textsc{a}C_\textsc{b} -
    \ii\hat\mu_\textsc{a}\rao\otimes C_\textsc{b}S_\textsc{a}\rof C_\textsc{a}C_\textsc{b} }\notag\\
    &\!\!+ \hat\mu_\textsc{b}\rbo\hat\mu_\textsc{b}\otimes
    \rr{\rao\otimes S_\textsc{b}C_\textsc{a}\rof C_\textsc{a}S_\textsc{b} +
    \hat\mu_\textsc{a}\rao\hat\mu_\textsc{a}\otimes S_\textsc{b}S_\textsc{a}\rof S_\textsc{a}S_\textsc{b} +
    \ii\rao\hat\mu_\textsc{a}\otimes S_\textsc{b}C_\textsc{a}\rof S_\textsc{a}S_\textsc{b} -
    \ii\hat\mu_\textsc{a}\rao\otimes S_\textsc{b}S_\textsc{a}\rof C_\textsc{a}S_\textsc{b} }\notag\\
    &\!\!+ \ii \rbo\hat\mu_\textsc{b}\otimes
    \rr{\rao\otimes C_\textsc{b}C_\textsc{a}\rof C_\textsc{a}S_\textsc{b} +
    \hat\mu_\textsc{a}\rao\hat\mu_\textsc{a}\otimes C_\textsc{b}S_\textsc{a}\rof S_\textsc{a}S_\textsc{b} +
    \ii\rao\hat\mu_\textsc{a}\otimes C_\textsc{b}C_\textsc{a}\rof S_\textsc{a}S_\textsc{b} -
    \ii\hat\mu_\textsc{a}\rao\otimes C_\textsc{b}S_\textsc{a}\rof C_\textsc{a}S_\textsc{b} }\notag\\
    &\!\!-\ii \hat\mu_\textsc{b}\rbo\otimes
    \rr{\rao\otimes S_\textsc{b}C_\textsc{a}\rof C_\textsc{a}C_\textsc{b} +
    \hat\mu_\textsc{a}\rao\hat\mu_\textsc{a}\otimes S_\textsc{b}S_\textsc{a}\rof S_\textsc{a}C_\textsc{b} +
    \ii\rao\hat\mu_\textsc{a}\otimes S_\textsc{b}C_\textsc{a}\rof S_\textsc{a}C_\textsc{b} -
    \ii\hat\mu_\textsc{a}\rao\otimes S_\textsc{b}S_\textsc{a}\rof C_\textsc{a}C_\textsc{b} }\,,
\end{align}
\end{widetext}
using the shorthand $C_j \equiv \cos\hat Y_j$ and $S_j\equiv \sin\hat Y_j$, and here it is understood that $\hat\mu_j\equiv \hat\mu_j(t_{j,0})$ in order to alleviate notation. For an algebraic state associated to $\rof$ (which defines the distinguished folium of normal states associated to some algebraic state $\mathcal{H}_\omega$, \textit{c.f.} Section~\ref{sec: AQFT} or \cite{hollands2017entanglement}) let us define another shorthand\footnote{This shorthand is slightly different from the definition given in \cite{tjoa2022channel} but simpler to use in terms of the ordering of the operators.}
\begin{equation}
    \begin{aligned}
    \gamma_{ijkl} &\coloneqq \omega(X^{(i)}_\textsc{a} X^{(j)}_\textsc{b}X^{(k)}_\textsc{b}X^{(l)}_\textsc{a}) \\
    &\equiv \tr(\rof X^{(i)}_\textsc{a} X^{(j)}_\textsc{b}X^{(k)}_\textsc{b}X^{(l)}_\textsc{a}) 
    \end{aligned}
\end{equation}
where $i,j,k,l = c,s$ for cosine and and sine respectively, e.g., $X^{(c)}_\textsc{a}\equiv \cos\hat Y_\textsc{a}$. By taking partial trace over $B$ and $\phi$, we get
\begin{align}
    \Phi(\rao) &= (\gamma_{cccc}+\gamma_{cssc}+\ii\beta (\gamma_{cscc}-\gamma_{ccsc}) )\rao  \notag\\
    &+ (\gamma_{sccs}+\gamma_{ssss}+\ii\beta (\gamma_{sscs}-\gamma_{scss}) ) \hat\mu_\textsc{a}\rao\hat\mu_\textsc{a}\notag\\
    &+ \ii(\gamma_{sccc}+\gamma_{sssc}+\ii\beta (\gamma_{sscc}-\gamma_{scsc}) ) \rao\hat\mu_\textsc{a}\notag\\
    &- \ii(\gamma_{cccs}+\gamma_{csss}+\ii\beta (\gamma_{cscs}-\gamma_{ccss}) ) \hat\mu_\textsc{a}\rao\,,
    \label{eq: channel-Bob}
\end{align}
where $\beta = \tr(\hat\mu_\textsc{b}\rbo)$. The coefficients can be computed straightforwardly for quasifree or Gaussian states using only properties of the Weyl algebra and Weyl relations \eqref{eq: Weyl-relations}. Note that as stated in Eq.~\eqref{eq: channel-Bob}, the result is valid for \textit{any} initial state of Alice and Bob, and furthermore it is also valid for any initial algebraic state of the field.

For convenience let us consider the field to be initially in a quasifree Hadamard state, which includes the vacuum state. This implies that $\gamma_{ijkl}=0$ if there are odd number of sines and cosines and we get
\begin{align}
    \Phi(\rao) &= (\gamma_{cccc}+\gamma_{cssc})\rao + (\gamma_{sccs}+\gamma_{ssss}) \hat\mu_\textsc{a}\rao\hat\mu_\textsc{a}\notag\\
    &+\beta (\gamma_{cscs}-\gamma_{ccss})\hat\mu_\textsc{a}\rao - \beta (\gamma_{sscc}-\gamma_{scsc}) \rao\hat\mu_\textsc{a}\,.
    \label{eq: channel-Bob-quasifree}
\end{align}
The remaining coefficients can be computed using the following identities proven in \cite{Tjoa2022fermi}:
\begin{lemma}
    We have the ``twisted'' product-to-sum formulae for Weyl algebra:
    \begin{subequations}
    \begin{align}
        2C_iC_j &= C_{i+j}e^{-\ii E_{ij}/2} + C_{i-j}e^{\ii E_{ij}/2} \,,\\
        -2S_iS_j &= C_{i+j}e^{-\ii E_{ij}/2} - C_{i-j}e^{\ii E_{ij}/2} \,,\\
        2C_iS_j &= S_{i+j}e^{-\ii E_{ij}/2} - S_{i-j}e^{\ii E_{ij}/2} \,,\\
        2S_iC_j &= S_{i+j}e^{-\ii E_{ij}/2} + S_{i-j}e^{\ii E_{ij}/2} \,,
    \end{align}
    \end{subequations}
    where $C_{i\pm j}\equiv \cos(\hat\phi(f_i\pm f_{j}))$, $S_{i\pm j}\equiv \sin(\hat\phi(f_i\pm f_{j}))$ and $E_{ij}\coloneqq E(f_i,f_j)$ is the smeared causal propagator. If $\supp(f_i)$ and $\supp(f_j)$ are spacelike-separated, we have $E_{ij}=0$ and these reduce to the standard product-to-sum formula in trigonometry for complex numbers (or for commuting operators).
\end{lemma}
\noindent The coefficients are given by 
\begin{subequations}
\begin{align}
    \gamma_{cccc}+\gamma_{cssc}&= \frac{1}{2}\rr{1+\nu_{\textsc{b}}\cos(2E_{\textsc{ab}})}\,,\\
    \gamma_{sccs}+\gamma_{ssss} &= \frac{1}{2}\rr{1 - \nu_{\textsc{b}}\cos(2E_{\textsc{ab}})}\,,\\
    \gamma_{cscs}-\gamma_{ccss} &= \gamma_{sscc}-\gamma_{scsc} =  \frac{\ii}{2}\nu_\textsc{b}\sin(2E_\textsc{ab})\,,
\end{align}
\end{subequations}
where $\nu_\textsc{b} = \omega(W(2Ef_\textsc{b})) = e^{-2\mathsf{W}(f_{\textsc{b}},f_{\textsc{b}})}$. Therefore, we can write
\begin{align}
    \Phi(\rao) &= \frac{1 + \nu_{\textsc{b}}\cos(2E_{\textsc{ab}})}{2}\rao + \frac{1 - \nu_{\textsc{b}}\cos(2E_{\textsc{ab}})}{2} \hat\mu_\textsc{a}\rao\hat\mu_\textsc{a}\notag\\
    &\hspace{0.4cm}+\frac{\nu_\textsc{b}\sin(2E_\textsc{ab})}{2} [\rao,\hat\mu_\textsc{a}]\,.
    \label{eq: channel-Bob-quasifree-2}
\end{align}
It is worth noting that if one wishes to be extremely strict with rigour and assumes that we cannot compute $\nu_\textsc{b}$ using definition of quasifree state because $f_\textsc{b}\not\in\CS$, the calculation of $\nu_\textsc{b}$ can be justified in canonical quantization using a more brute-force approach along the lines of \cite{perche2022spacetime}.

This channel $\Phi$ has been shown to have channel capacity equal to \cite{tjoa2022channel}
\begin{align}
    C(\Phi) &= H\Bigr(\frac{1}{2}+\frac{\nu_\textsc{b}}{2}\left|\cos2E(f_\textsc{a},f_\textsc{b})\right|\Bigr)-H\Bigr(\frac{1}{2}+\frac{\nu_\textsc{B}}{2}\Bigr)\,,
\end{align}
where $H$ is the binary Shannon entropy. Crucially, the channel capacity is zero if Alice and Bob's qubits are spacelike-separated, as we expect from first-principle relativistic communication. Another way to see this is that relativistic causality is encoded in the coefficients of $\Phi$: if Alice and Bob are spacelike separated, the smeared causal propagator $E(f_{\textsc{a}},f_{\textsc{b}}) = 0$ and Bob's final state after interaction is purely due to local noise coming from its interaction with the field (since it only depends on $f_\textsc{b}$). In \cite{tjoa2022channel} it was also shown that the maximum classical channel capacity can be attained with suitable tuning of the detector parameters --- in particular, it is generically the case that Alice's coupling must be much stronger than Bob for the capacity to be maximized, and Bob just needs to pick a pure state $\rbo=\ketbra{g_\textsc{b}}{g_\textsc{b}}$. Therefore, Alice can communicate her classical message perfectly to Bob. Once Bob performs decoding operation to infer Alice's message $m$, Bob can then perform the corresponding unitary in Eq.~\eqref{eq: control-unitary}. 

This result is quite general since the channel is completely specified by the properties of Bob's detector, initial state, and the expectation value of the quantum field observables in arbitrary curved spacetime and is also independent of the GNS representation we consider so long as the algebraic state is quasifree.

\section{Possible extensions and variants}
\label{sec: extension}

In this section we briefly comment on how our protocol can be extended to fit the sort of setups considered in the past literature.

\subsection{Including environmental effects and states of motion}

As we mentioned in the preceding section, the earlier ``relativistic teleportation'' protocol is not strictly speaking a relativistic protocol because what is relativistic is the ``environment'' and the states of motion of the qubits in the kinematical sector $\mathcal{H}_\textsc{kin}$. Recall that in \cite{Landulfo2009suddendeath,Alsing2003teleport}, what was being considered is the impact of quantum field environment interacting locally with Alice and Bob's detectors living in $\mathcal{H}_{A}$ and $\mathcal{H}_{B}$ (while leaving $\mathcal{H}_{A_1}$ intact). Indeed this part of the analysis is relativistic in the sense that the quantum field constitutes a relativistic bath, while the detectors are allowed to move along relativistic trajectories consistent with special relativity. We argued that what is necessary in this case is to make use of the quantum field as part of the communication (sub)protocol. 

Given the construction in Section~\ref{sec: teleportation-protocol}, it is now clearer how to include the setup considered in \cite{Alsing2003teleport,Landulfo2009suddendeath}. The idea is to have the same quantum field interact with all five qubit sectors. For convenience we write down the six Hilbert spaces from Eq.~\eqref{eq: six-hilbert-spaces}:
\begin{align*}
    \mathcal{H} &= \underbrace{\mathcal{H}_{A_1}\otimes\mathcal{H}_{A}\otimes\mathcal{H}_{B} }_{\text{kinematical}}\otimes\,\underbrace{\mathcal{H}_{A_2}\otimes\mathcal{H}_{B_1}\otimes\mathcal{H}_\phi}_{\text{communication }}\,.
\end{align*}
This time round, however, we will need at least four qubits $A,B,A_2,B_1$ to interact with the quantum field $\phi$ (while protecting Alice's other half of the Bell state\footnote{We could easily accommodate all five qubits interacting but we do not do so here for simplicity.}). All we need is to consider the same UDW detector coupling in \eqref{eq: hamiltonian-four-form} for all four qubits: the total unitary is now sum of four Hamiltonian four-forms:
\begin{align}
    U = \mathcal{T}\exp\left[-\ii\sum_{j}\int_\M\!\!\!\!\dd V\,h_{j,I}(\sx)\right]\,,
\end{align}
where $j = A,B,A_2,B_1$. The spacetime smearings for each qubit should be chosen so that $f_{A}$ and $f_{A_2}$ are spacelike separated (e.g., at the same Cauchy slice $\Sigma_t$ but separated by finite but small proper distance), similarly $f_{B},f_{B_1}$ are also spacelike (e.g., also at the same Cauchy slice $\Sigma_{t'}$ but separated by finite but small proper distance). Then Bob carries \textit{both} qubits $B$ and $B_1$ along some relativistic trajectories in a common moving laboratory and we can perform the same type of calculations along the spirit of \cite{Alsing2003teleport,Landulfo2009suddendeath} to account for the effects coming from states of motion\footnote{Of course, as before, one has to be careful about time ordering in delta coupling approach if motion is not stationary; for practical purposes we require that $t(\supp f_\textsc{a})\leq t(\supp f_\textsc{b})$, a condition analogous to that imposed by \cite{Landulfo2016magnus1}.}.

Since the communication part (which requires maximum channel capacity) only involves $A_2$ and $B_1$, the dynamics of detectors $A$ and $B$ can be evaluated perturbatively so we do not need to make any restrictions to the type of interactions or detectors, beyond the assumption that the coupling with the field is weak (to justify perturbative calculation). If we wish to study nonperturbatively how non-trivial states of motion can impact the teleportation protocol, then we can consider the moving detectors $A$ and $B$ to be gapless (as done in \cite{Landulfo2016magnus1}) since the delta-coupled detector cannot really capture the impact of the states of motion beyond the fact that the spatial profile gets distorted asymmetrically, e.g., when one considers detector $B$'s centre of mass to be uniformly accelerated (see, for instance, \cite{Tales2022FNC} for how to capture accelerated motion of a finite-sized detector in Fermi normal coordinates). 

One complication with including two extra qubits' states of motion is that it is no longer obvious the channel capacity can be calculated as easily, since the quantum channel $\Phi$ will get modified by additional interactions coming from the detectors in the kinematical sector $A$ and $B$. To get a sense of what has changed, note that having four detectors couple to the field would mean that the coefficients of the channel would not in general depend only on products of four operators with two distinct smearing functions (since $\gamma_{ijkl}$ depends on $f_{A_2}$ and $f_{B_1}$). This suggests that the calculation's complexity parallels that of having two detectors each coupling to the field \textit{twice}, as done in \cite{Simidzija2018nogo} in the context of entanglement harvesting, so the new sets of coefficients will involve eight indices $\tilde{\gamma}_{ijklmnop}$ depending on four spacetime smearings $f_{A},f_{B},f_{A_2},f_{B_1}$. Since we know from \cite{Simidzija2020transmit,jonsson2018transmitting} that perturbative coupling can only influence channel capacity by perturbatively small amount, we expect that if the kinematical sector's detectors $A,B$ interact weakly with the field, then the channel capacity will still be close to maximal. However, simply from the fact that detector-field interaction introduces some noise, we know that the teleportation fidelity by including the kinematical sector's qubits $A,B$ will be worse than the situation we considered in Section~\ref{sec: teleportation-protocol}.

\subsection{Another possible form of classical communication?}

One other (na\"ive) possibility is to consider a different way of sending the classical information to Bob, making use of the coupling with the field. For example, one considers applying the UDW coupling of the form
\begin{subequations}
\begin{align}
    00 \to H^m_I(t) &=     \openone \otimes \openone \,,\\
    01 \to H^m_I(t) &= \hat\sigma^z \otimes \hat\phi(f) \,,\\
    10 \to H^m_I(t) &= \hat\sigma^x \otimes \hat\pi(f) \,,\\
    11 \to H^m_I(t) &= \hat\sigma^y \otimes \hat\pi(f)\hat\phi(f) \,.
\end{align}
\end{subequations}
Here the 3-smeared field operators are
\begin{align}
    \hat\phi(f) &= \int_{\Sigma_t}\dd^3\bx\sqrt{h}\,\hat\phi(t,\bx)f(t,\bx)\,,
    \\
    \hat\pi(f) &= \int_{\Sigma_t}\dd^3\bx\sqrt{h}\,t^a\nabla_a\phi(t,\bx)f(t,\bx)\,,
\end{align}
with $\pi(\sx) = \sqrt{h} t^a\nabla_a\phi$. The field operator $\hat\phi(f)$ in this case can be obtained from the four-smeared field operator $\hat\phi(f)$ in the sense of Section~\ref{sec: AQFT} but for delta coupling case we can technically regard it as 3-smeared operator at some fixed time slice\footnote{Note that the interpretation of the Weyl algebra differs somewhat from Section~\ref{sec: AQFT} if we work with CCR representation over a Cauchy surface, see \cite{dimock1980algebras,KayWald1991theorems}.

}.

For this proposal, because the field helps in distinguishing messages, in principle Alice only needs one auxiliary detector to do this; part of the classical message will be encoded into the vacuum fluctuations and Bob's measurement would differ because each interaction Hamiltonian induces different kinds of channel $\mathcal{E}_m$ associated to Hamiltonian $H_{I}^m(t)$. In other words, Bob's measurement is associated to the following: 
\begin{align}
    \sigma^{A_2B_1}_m &= \ketbra{m}{m}\otimes \tr(E_m\Phi_m(\rho_{0}^{A_1}))\ketbra{m}{m}\,.
\end{align}
where now each channel $\Phi_m$ is associated to each distinct message $m\in M$. 

In more detail, the key observation is that the channel $\Phi_m$ ($m=00,01,10,11$) takes exactly the same functional form as $\Phi$ defined in Eq.~\eqref{eq: channel-Bob-quasifree-2}: that is,
\begin{align}
    \Phi_m(\rao) &= \frac{1 + \nu_{\textsc{b}}^m\cos(2E_{\textsc{ab}})}{2}\rao +\frac{\nu_\textsc{b}^m\sin(2E_\textsc{ab})}{2} [\rao,\hat\mu_{\textsc{a}}^m]\notag\\
    &\hspace{0.4cm}+ \frac{1 - \nu_{\textsc{b}}^m\cos(2E_{\textsc{ab}})}{2} \hat\mu_\textsc{a}^m\rao\hat\mu_\textsc{a}^m\,,
    \label{eq: channel-Bob-quasifree-mod}
\end{align}
where $\hat\mu_\textsc{a}^m = \openone\,,\hat\sigma^x\,,\hat\sigma^y\,,\hat\sigma^z$ depending on the values of $m$, and similarly
\begin{align}
    \nu_\textsc{b}^m &= \omega(e^{\ii \hat O_m(f)})\,,
\end{align}
where $\hat O_m(f) = \openone\,, \hat\phi(f)\,,\hat\pi(f)\,,\hat\pi(f)\hat\phi(f)$ depending on which value of $m$. Alice should also pick her initial state to be such that it does not commute with her monopole operator (otherwise the result will not be distinguishable from $m=00$).

The success of this approach relies on the fact that the monopole operators $\hat\mu_\textsc{a}^m$ are distinct (and orthogonal in Hilbert-Schmidt inner product) and the fact that $\nu_\textsc{b}^m$ can be  different. We have
\begin{align}
    \nu_\textsc{b}^{00}&= 1\,,\quad \nu_\textsc{b}^{01}= e^{-\mathsf{W}(f,f)/2}\,,
\end{align}
while for $\nu_\textsc{b}^{10},\nu_\textsc{b}^{11}$ it is more complicated because $\hat\pi(f)$ only makes sense as 3-smeared operator (while $\hat\phi(f)$ is 4-smeared operator\footnote{Again, in delta-coupled regime we are sort of abusing the rigour, since $f\not\in\CS$. So one has to think of delta-coupled regime as compactly supported smooth function which is very tightly supported along the time direction. If one finds this approach not palatable but still wishes for non-perturbative results, one can always pass to gapless detector regime where this issue does not arise (but detector energy levels are degenerate and they have no internal dynamics).}). However, all we need to know is that they are all distinct, which follows from the fact that $e^{\ii \hat O_m(f)}$ is a (product of) displacement operator(s). Even for $\nu_{\textsc{b}}^{11}$, we have from Baker-Campbell-Hausdorff formula
\begin{align}
    e^{\ii \hat O_{11}(f)} = e^{\ii \hat\phi(f)}e^{\ii \hat\pi(f)}e^{-\ii \theta \openone}
\end{align}
where $\theta$ is some constant phase. This is a direct consequence of (smeared) equal-time commutation relation at fixed Cauchy slice \cite{KayWald1991theorems,dimock1980algebras}. In flat space, this can be checked explicitly (see \cite{Simidzija2020transmit} for an analogous calculation in entanglement harvesting). All Bob needs to make sure, then, is to pick a pure state $\rbo$ that does not commute with any of the Pauli matrices. Suppose $[\rbo,\hat\sigma^y]=0$, then $\rbo$ will be a fixed point of $\Phi_{11}$, and so Bob is unable to distinguish $\Phi_{00}$ from $\Phi_{11}$ regardless of his choice of POVMs since $\Phi_{00}=\Phi_{11}$ in that case. Note that since the channel $\Phi_m$ is practically identical to the previous approach analogous to \cite{tjoa2022channel}, we expect that the channel capacity can be made equally large by suitable adjustment of the detector parameters and choice of states for embedding the classical information.

One complication with this approach is that if Alice wishes to send $00$, then it amounts to doing nothing so both Alice and Bob have to agree in advance that if at the prescribed time Bob receives nothing from Alice via the quantum channel, then Bob does nothing as well and take $\ket{\zeta_{00}}$ to be the actual state Alice teleported. This is different from if Bob is causally disconnected from Alice, where Bob must average over all possible outcomes according to the protocol, and it is not always clear whether Bob can make this distinction. One possible way out of this issue is to instead consider more general qudits, but the detector-field interaction must be given in terms of clock-and-shift matrices analogous to the approach adopted in \cite{qet2016qudit}. This allows more non-trivial operators for Alice to use to encode her classical information, since ideally Alice would want four distinct (orthogonal) operators that are not equal to the identity operator.

\section{Conclusion}

In this work we close the gap in relativistic quantum information (RQI) by proposing a genuine relativistic quantum teleportation protocol, in contrast to the past proposals given in \cite{Alsing2003teleport,Landulfo2009suddendeath}. Our protocol is based on the minimal requirement that at the fundamental level, all the {relevant} components for the protocol must respect relativistic principles and the information and qubits are physically accessible. While our protocol is unlikely to be useful from practical or experimental standpoints, our goal is to show that we can have relativistically consistent quantum teleportation protocol that fully respects causality \textit{by construction}. The practical implementation can then be viewed as a ``coarse-grained''  protocol where relativity is no longer necessary for the user to be aware of and be treated as a black box component of the protocol. 

The relativistic teleportation protocol we propose here is also based on the UDW detector model but generalized to arbitrary globally hyperbolic Lorentzian manifold $(\M,g_{ab})$, using the sort of non-perturbative setups considered in \cite{Landulfo2016magnus1,tjoa2022channel}. The non-perturbative method is needed to ensure that the classical channel capacity is not perturbatively small. We stress again that to accommodate fully relativistic quantum teleportation protocol, what needs to be changed is the classical communication part and less on the fact that the ``noisy environment'' is a relativistic quantum field. If we simply assume that the LOCC are perfect \textit{somehow}, then the quantum teleportation protocol (or really, any of the standard quantum protocols in quantum information theory) is completely independent of whether relativity theory holds, or even whether spacetime exists at all (see, e.g., \cite{vilasini2022embedding,Paunkovic2020causalorders} and refs. therein).

This work has several natural extensions for future work. First, since our goal is to set this up to make sure that the relativistic component is obvious from first principles, we have not made any attempt to really control or evaluate the one-shot channel capacity in relativistic settings. Clearly, since the hypothesis testing channel capacity in Section~\ref{sec: teleportation-protocol} is a function of the relativistic channel, we expect that spacelike separated detectors have zero one-shot channel capacity for any $\epsilon\in [0,1)$. It would be desirable to obtain more explicit expression for this, in the same way the entanglement-breaking property allows for the asymptotic Holevo information to be computed. The one-shot analysis is also conceptually useful because ``repeated use of the channel'' is not easy to think about in relativistic settings --- we cannot literally ``reset'' the background spacetime (though this approximation is sensible, since this is how particle physics experiments make sense). There may be ways to account for this more simply.

Second, as per constructions in \cite{Landulfo2016magnus1,tjoa2022channel}, we are still considering spherically symmetric broadcast using scalar fields. For classical capacity, this is not a problem but for quantum capacity this is known to be detrimental in general without having Bob to be delocalized over large regions \cite{Simidzija2020transmit}. Analyses of ``directional'' transmission of classical information and classical propagation of information using non-scalar fields are of independent interest. Also, note that in \cite{Simidzija2020transmit} the authors used two detectors coupled to the field locally \textit{twice} in some appropriate sense (though specialized to flat spacetimes) and this allowed for transmission of \textit{qubits} with arbitrarily high quantum channel capacity. This could be used to replace the CC component since one can then use this to transmit classical information as well. However, intuitively we expect that typically quantum channels that can transmit quantum information are ``costlier'' and more difficult to work with than those that can only transmit classical capacity, so it would be nice to see if our proposals are really ``simpler'' than using the kind of settings in \cite{Simidzija2020transmit}.

Third, one of the interesting applications of quantum teleportation is the so-called \textit{gate-based teleportation} \cite{gottesman1999demonstrating,jozsa2006introduction}. This allows for universal quantum computation using only single-qubit gates, Bell measurements, and also tripartite entangled Greenberger-Horne-Zeilinger (GHZ) states. There is also a variant of teleportation called \textit{remote state preparation} (RSP) \cite{Bennett2001remote,Bennett2002erratum}. Here the RSP protocol differs from teleportation in that Alice \textit{knows} what state to send to Bob, so the resources required to perform this task is fewer. It is of independent interest to see if relativistic version of teleportation protocol can be adapted for these applications and see whether there is any additional features or insights to be gained from them (for RSP we expect, at least, that it will not be too different). 

Last but not least, with the recent constructions of Fewster-Verch framework for local measurement theory \cite{fewster2020quantum}, it is perhaps natural to see if there is a way to fit teleportation into such a framework where the probe is also a quantum field. While the physicality of the measurement schemes, the general approach and the comparison with the UDW model are still part of an ongoing debate, from a mathematical standpoint we find it a worthwhile line of investigation simply because the nature of what is being teleported is not obvious at all. Since we know that we cannot teleport a single field mode as done in \cite{Alsing2003teleport}, it is desirable to even know whether it makes sense to talk about teleportation when \textit{all} components are purely quantum fields. We believe there may be a way to do this, based on two separate ideas: (1) that the split property of the quantum fields allows us to think of embedding ``Cbits'' into spacetime regions (see, e.g., \cite{hollands2017entanglement}), and (2) commonly used finite-dimensional concepts such as entanglement distillation and positive partial transpose make sense even in quantum field theory (see, e.g., \cite{Verch2005distill}). As is usual for a typical theoretical physicist, we leave these (very!) interesting lines of investigations for future work.

\section*{Acknowledgments}

E.T. thanks Chris Fewster and Atsushi Higuchi for useful discussions that resulted in this work, and also T. thanks Tales Rick Perche for clarifying the use of Fermi normal coordinates.  E. T. acknowledges that this work is made possible from the funding through both of his supervisors  Robert B. Mann and Eduardo Mart\'in-Mart\'inez. This work is supported in part by the Natural Sciences and Engineering Research Council of Canada (NSERC).  The funding support from Eduardo Mart\'in-Mart\'inez is through the Ontario Early Research Award and NSERC Discovery program. E. T. would also like to thank both supervisors for providing the opportunity to explore some research ideas independently, through which this work becomes possible.


\appendix 

\section{Condensed review of scalar QFT in curved spacetimes}
\label{sec: AQFT}

In order to ensure that this work is self-contained, in this Appendix we very briefly review the algebraic framework for quantization of a real scalar field in arbitrary (globally hyperbolic) curved spacetimes, which is a condensed review of \cite{Tjoa2022fermi}. Readers can find accessible introduction to $*$-algebras and $C^*$-algebras for QFT applications in \cite{fewster2020algebraic,hollands2017entanglement,Khavkhine2015AQFT,KayWald1991theorems}. Readers who are more interested in the bird's eye view of the relativistic generalization are invited to \tcb{read Section~\ref{sec: teleportation-protocol} and consult this Appendix as the need arises.}

\subsection{Algebra of observables}

For this work, we restrict our attention to a real scalar field $\phi$ in (3+1)-dimensional globally hyperbolic Lorentzian spacetime $(\mathcal{M},g_{ab})$, whose equation of motion is given by
\begin{align}
     P\phi = 0\,,\quad  P = \nabla_a\nabla^a - m^2  - \xi R\,,
     \label{eq: KGE}
\end{align}
where $\xi \geq 0$, $R$ is the Ricci scalar and  $\nabla$ is the Levi-Civita connection with respect to $g_{ab}$. Since it is globally hyperbolic, it admits a foliation  $\M\cong \R\times \Sigma$ where $\Sigma$ is a Cauchy surface and there is a good notion of global time ordering (we can speak of ``constant-time surfaces'').

The big picture of the algebraic framework for free scalar field quantization is as follows. First, we need to construct the \textit{algebra of observables} $\A(\M)$ for the field theory as well as quantum states on which $\A(\M)$ acts. The building blocks of the quantum theory are made out of solutions to \eqref{eq: KGE}, which in turn can be constructed out of certain Green's functions. We also need to provide an implementation of dynamics and  \textit{canonical commutation relations} (CCR). Finally, we also need to provide a family of quantum states without reference to any Hilbert space structure: in QFT there are many unitarily inequivalent Hilbert space representations. A physically reasonable choice of such set of states are known as \textit{Hadamard states} \cite{KayWald1991theorems,Radzikowski1996microlocal}, which respects local flatness property and certain regularity conditions.

First, we need to construct solutions to build the algebra of observables. Let $f\in \CS$ be a smooth compactly supported test function on $\M$. We consider the \textit{retarded and advanced propagators} $E^\pm\equiv E^\pm(\sx,\sy)$ associated to the Klein-Gordon operator $P$, defined to be the Green's functions such that they solve the inhomogeneous wave equation. That is, we have $P(E^\pm f) = f$, with
\begin{align}
    E^\pm f\equiv (E^\pm f)(\sx) \coloneqq \int \dd V'\, E^\pm (\sx,\sx')f(\sx') \,,
\end{align}
where $\dd V' = \dd^4\sx'\sqrt{-g}$ is the invariant volume element. The \textit{causal propagator} is defined to be the advanced-minus-retarded propagator $E=E^--E^+$. The key fact we need is that if $O$ is an open neighbourhood of some Cauchy surface $\Sigma$ and $\varphi \in \Sol_\R(\M)$ is any real solution to Eq.~\eqref{eq: KGE} with compact Cauchy data, then there exists $f\in \CS$ with $\supp(f)\subset O$ such that $\varphi=Ef$ \cite{Khavkhine2015AQFT}.

The quantization of the real scalar field theory $\phi$ in AQFT is regarded as an $\R$-linear mapping from the space of smooth compactly supported test functions to a unital $*$-algebra $\A(\M)$
\begin{align}
    \hat\phi: C^\infty_0(\mathcal{M})&\to \A(\M)\,,\quad f\mapsto \hat\phi(f)\,,
\end{align}
together with the following requirements:
\begin{enumerate}[leftmargin=*,label=(\alph*)]
    \item (\textit{Hermiticity}) $\hat\phi(f)^\dag = \hat\phi(f)$ for all $f\in \CS$;
    \item (\textit{Klein-Gordon}) $\hat\phi(Pf) = 0$ for all $f\in \CS$;
    \item (\textit{Canonical commutation relations}  (CCR)) $[\hat\phi(f),\hat\phi(g)] = \ii E(f,g)\openone $ for all $f,g\in \CS$, where $E(f,g)$ is the smeared causal propagator
    \begin{align}
        E(f,g)\coloneqq \int \dd V f(\sx) (Eg)(\sx)\,.
    \end{align}
    \item (\textit{Time slice axiom})  $\A(\M)$ is generated by the unit element $\openone$ and the smeared field operators $\hat\phi(f)$ for all $f\in \CS$ with $\supp(f)\subset O$, where $O$ a fixed open neighbourhood of some Cauchy slice $\Sigma$.
\end{enumerate}
$\A(\M)$ is called the \textit{algebra of observables} for the scalar field theory and the \textit{smeared} field operator reads
\begin{align}\label{eq: ordinary smearing}
    \hat\phi(f) = \int \dd V\hat\phi(\sx)f(\sx)\,.
\end{align}
What we usually use in canonical quantization, namely (unsmeared) field operator $\hat\phi(\sx)$, is formally an operator-valued distribution.  

Next, the vector space of solutions $\Sol_\R(\M)$ can be equipped with a symplectic form $\sigma:\Sol_\R(\M)\times\Sol_\R(\M)\to \R$, defined as
\begin{align}
    \sigma(\phi_1,\phi_2) \coloneqq \int_{\Sigma_t}\!\! {\dd\Sigma^a}\,\Bigr[\phi_{{1}}\nabla_a\phi_{{2}} - \phi_{{2}}\nabla_a\phi_{{1}}\Bigr]\,,
    \label{eq: symplectic form}
\end{align}
where $\dd \Sigma^a = -t^a \dd\Sigma$, $-t^a$ is the inward-directed unit normal to the Cauchy surface $\Sigma_t$, and $\dd\Sigma = \sqrt{h}\,\dd^3\bx$ is the induced volume form on $\Sigma_t$ \cite{Poisson:2009pwt,wald2010general}. For our purposes, the symplectic form \eqref{eq: symplectic form} is mainly to relate to the usual canonical quantization procedure, since we will need to have \textit{Klein-Gordon inner product} for the one-particle Hilbert space (sometimes called ``positive-frequency Hilbert space'').

In this work, as is standard we will work with the ``exponentiated version''of $\hat\phi(f)$ which forms a \textit{Weyl algebra} $\W(\M)$, which is a unital $C^*$-algebra generated by elements that formally take the form 
\begin{align}
    W(Ef) \equiv 
    {e^{\ii\hat\phi(f)}}\,,\quad f\in \CS\,,    \label{eq: Weyl-generator}
\end{align}
obeying \textit{Weyl relations}:
\begin{equation}
    \begin{aligned}
    W(Ef)^\dagger &= W(-Ef)\,,\\
    W(E (Pf) ) &= \openone\,,\\
    W(Ef)W(Eg) &= e^{-\frac{\ii}{2}E(f,g)} W(E(f+g))
    \end{aligned}
    \label{eq: Weyl-relations}
\end{equation}
where $f,g\in \CS$. The Weyl algebra is more useful here since the elements are bounded operators and we do not have to deal with functional-analytic domain issues associated to $\A(\M)$. Causal behaviour associated to two disjoint regions $f,g$ is encoded in the third Weyl relations due to the appearance of causal propagator $E(f,g)$.

\subsection{Algebraic states and quasifree states}

An \textit{algebraic state} is defined to be a $\C$-linear functional $\omega:\W(\M)\to \C$ (similarly for $\A(\M)$) such that 
\begin{align}
    \omega(\openone) = 1\,,\quad  \omega(A^\dagger A)\geq 0\quad \forall A\in \W(\M)\,.
    \label{eq: algebraic-state}
\end{align}
In words, the algebraic state is essentially a map from observables to expectation values. We say that the state $\omega$ is pure if it cannot be written as $\omega= \alpha \omega_1 + (1-\alpha)\omega_2$ for any $\alpha\in (0,1)$ and any two algebraic states $\omega_1,\omega_2$ and it is mixed otherwise. 

The \textit{Gelfand-Naimark-Segal (GNS) reconstruction theorem} \cite{wald1994quantum,Khavkhine2015AQFT,fewster2020algebraic} relates the algebraic approach to the canonical approach. The theorem says that given $\W(\M)$ and $\omega$, we get a \textit{GNS triple} $(\mathcal{H}_\omega, \pi_\omega,{\ket{\Omega_\omega}})$, where $\pi_\omega: \mathcal{\W(\M)}\to {\mathcal{B}(\mathcal{H}_\omega)}$ is a Hilbert space representation with respect to $\omega$. Any algebraic state $\omega$ is realized as a \textit{vector state} $\ket{\Omega_\omega}\in\mathcal{H}_\omega$  in its own GNS representation, and observables  $A \in \W(\M)$ are realized as bounded operators $\hat A\coloneqq \pi_\omega(A)\in \mathcal{B}(\mathcal{H}_\omega)$. The expectation values will then take the familiar form $ \omega(A) = \braket{\Omega_\omega|\hat A|\Omega_\omega}$. The GNS theorem allows us to only deal with specific representations (out of infinitely many inequivalent ones) ``at the end''.

We can essentially characterize the scalar QFT by computing the \textit{$n$-point correlation functions} (also known as the Wightman $n$-point functions), given by
\begin{align}
    \mathsf{W}(f_1,...,f_n)\coloneqq \omega(\hat\phi(f_1)...\hat\phi(f_n))\,,
    \label{eq: n-point-functions}
\end{align}
where $f_j\in \CS$. Although the RHS seems to involve the algebra $\A(\M)$, the GNS representation of the Weyl algebra $\W(\M)$ lets us calculate these correlators using derivatives: for example, we have the smeared Wightman two-point function
\begin{align}\label{eq: Wightman-formal-bulk}
    &\mathsf{W}(f,g) \equiv -\frac{\partial^2}{\partial s\partial t}\Bigg|_{s,t=0}\!\!\!\!\!\!\!\!\omega(e^{\ii\hat\phi(sf)}e^{\ii\hat\phi(tg)})
\end{align}
where the RHS should be viewed in the GNS representation of $\W(\M)$.

It is generally accepted that any physically reasonable states should be Hadamard states \cite{KayWald1991theorems,Radzikowski1996microlocal}. The specific features of Hadamard states need not concern us in this work: all we need is that there are nice subsets of Hadamard states that are \textit{quasifree}: these have vanishing odd-point correlation functions and all higher even-point functions can be written as in terms of just two-point functions. We reserve the phrase \textit{Gaussian states} to non-quasifree states where all higher-point functions only depend on one- and two-point functions. Examples of quasifree states include vacuum, squeezed vacuum, and thermal states; examples of non-quasifree Gaussian states include coherent states and squeezed coherent states. 

In fact, for our purposes we can simply \textit{define} a quasifree state $\omega$ as those given by
\begin{align}
    \omega(W(Ef)) =   e^{-{\frac{1}{2}}\mathsf{W}(f,f)}\,.
    \label{eq: quasifree-definition}
\end{align}
(see, e.g., \cite{KayWald1991theorems,Khavkhine2015AQFT} for more details). In situations with enough symmetries, the RHS can be computed in closed form. The most important one for us is the vacuum state $\omega_0$, whose (unsmeared) vacuum Wightman function takes a familiar form
\begin{align}
    \mathsf{W}_0(\sx,\sy) &= \int \dd^3\bk\, u^{\phantom{*}}_\bk(\sx) u^*_\bk(\sy)\,,
\end{align}
where $u_\bk(\sx)$ are ``positive-frequency'' modes of Klein-Gordon operator $P$ normalized with respect to Klein-Gordon inner product $(\phi_1,\phi_2)_\textsc{kg}\coloneqq \ii\sigma(\phi_1^*,\phi_2)$, where $\phi_j\in \Sol_\C(\M)$ are complexified solutions to Eq.~\eqref{eq: KGE} (\textit{c.f.} \cite{birrell1984quantum}). The smeared version is of course given by
\begin{align}
    \mathsf{W}_0(f,f) = \int \dd V\,\dd V' f(\sx)f(\sy)\mathsf{W}_0(\sx,\sy)\,.
    \label{eq: Wightman-double-smeared}
\end{align}
In most situations, the relevant states of interest gives Wightman correlators that are related to the vacuum correlators in the sense that they take the form
\begin{align}
    \mathsf{W}(f,g) &= \mathsf{W}_0(f,g) + \Delta\mathsf{W}(f,g)\,,
    \label{eq: decompose-wightman}
\end{align}
where $\Delta\mathsf{W}(f,g)$ is an extra term that depends on geometry and the state \cite{DeWitt1960radiation,wald1994quantum,KayWald1991theorems}. In flat spacetimes, we can actually calculate this extra term straightforwardly for well-known Gaussian states such as thermal states, coherent and squeezed states (see, e.g., \cite{simidzija2018harvesting,Weldon2000thermal}). Note that the well-known inequivalence of Rindler and Minkowski vacua refers to distinct vacuum component (i.e., they cannot be related by Eq.~\eqref{eq: decompose-wightman} in all of $\M$), since the Rindler vacuum is really associated to one ``wedge'' of Minkowski spacetime.

\bibliography{teleport-ref}

\begin{thebibliography}{77}%
\makeatletter
\providecommand \@ifxundefined [1]{%
 \@ifx{#1\undefined}
}%
\providecommand \@ifnum [1]{%
 \ifnum #1\expandafter \@firstoftwo
 \else \expandafter \@secondoftwo
 \fi
}%
\providecommand \@ifx [1]{%
 \ifx #1\expandafter \@firstoftwo
 \else \expandafter \@secondoftwo
 \fi
}%
\providecommand \natexlab [1]{#1}%
\providecommand \enquote  [1]{``#1''}%
\providecommand \bibnamefont  [1]{#1}%
\providecommand \bibfnamefont [1]{#1}%
\providecommand \citenamefont [1]{#1}%
\providecommand \href@noop [0]{\@secondoftwo}%
\providecommand \href [0]{\begingroup \@sanitize@url \@href}%
\providecommand \@href[1]{\@@startlink{#1}\@@href}%
\providecommand \@@href[1]{\endgroup#1\@@endlink}%
\providecommand \@sanitize@url [0]{\catcode `\\12\catcode `\$12\catcode
  `\&12\catcode `\#12\catcode `\^12\catcode `\_12\catcode `\%12\relax}%
\providecommand \@@startlink[1]{}%
\providecommand \@@endlink[0]{}%
\providecommand \url  [0]{\begingroup\@sanitize@url \@url }%
\providecommand \@url [1]{\endgroup\@href {#1}{\urlprefix }}%
\providecommand \urlprefix  [0]{URL }%
\providecommand \Eprint [0]{\href }%
\providecommand \doibase [0]{https://doi.org/}%
\providecommand \selectlanguage [0]{\@gobble}%
\providecommand \bibinfo  [0]{\@secondoftwo}%
\providecommand \bibfield  [0]{\@secondoftwo}%
\providecommand \translation [1]{[#1]}%
\providecommand \BibitemOpen [0]{}%
\providecommand \bibitemStop [0]{}%
\providecommand \bibitemNoStop [0]{.\EOS\space}%
\providecommand \EOS [0]{\spacefactor3000\relax}%
\providecommand \BibitemShut  [1]{\csname bibitem#1\endcsname}%
\let\auto@bib@innerbib\@empty
\bibitem [{\citenamefont {Bennett}\ \emph {et~al.}(1993)\citenamefont
  {Bennett}, \citenamefont {Brassard}, \citenamefont {Cr\'epeau}, \citenamefont
  {Jozsa}, \citenamefont {Peres},\ and\ \citenamefont
  {Wootters}}]{Bennett1993teleport}%
  \BibitemOpen
  \bibfield  {author} {\bibinfo {author} {\bibfnamefont {C.~H.}\ \bibnamefont
  {Bennett}}, \bibinfo {author} {\bibfnamefont {G.}~\bibnamefont {Brassard}},
  \bibinfo {author} {\bibfnamefont {C.}~\bibnamefont {Cr\'epeau}}, \bibinfo
  {author} {\bibfnamefont {R.}~\bibnamefont {Jozsa}}, \bibinfo {author}
  {\bibfnamefont {A.}~\bibnamefont {Peres}},\ and\ \bibinfo {author}
  {\bibfnamefont {W.~K.}\ \bibnamefont {Wootters}},\ }\bibfield  {title}
  {\bibinfo {title} {Teleporting an unknown quantum state via dual classical
  and einstein-podolsky-rosen channels},\ }\href
  {https://doi.org/10.1103/PhysRevLett.70.1895} {\bibfield  {journal} {\bibinfo
   {journal} {Phys. Rev. Lett.}\ }\textbf {\bibinfo {volume} {70}},\ \bibinfo
  {pages} {1895} (\bibinfo {year} {1993})}\BibitemShut {NoStop}%
\bibitem [{\citenamefont {Bouwmeester}\ \emph {et~al.}(1997)\citenamefont
  {Bouwmeester}, \citenamefont {Pan}, \citenamefont {Mattle}, \citenamefont
  {Eibl}, \citenamefont {Weinfurter},\ and\ \citenamefont
  {Zeilinger}}]{bouwmeester1997experimental}%
  \BibitemOpen
  \bibfield  {author} {\bibinfo {author} {\bibfnamefont {D.}~\bibnamefont
  {Bouwmeester}}, \bibinfo {author} {\bibfnamefont {J.-W.}\ \bibnamefont
  {Pan}}, \bibinfo {author} {\bibfnamefont {K.}~\bibnamefont {Mattle}},
  \bibinfo {author} {\bibfnamefont {M.}~\bibnamefont {Eibl}}, \bibinfo {author}
  {\bibfnamefont {H.}~\bibnamefont {Weinfurter}},\ and\ \bibinfo {author}
  {\bibfnamefont {A.}~\bibnamefont {Zeilinger}},\ }\bibfield  {title} {\bibinfo
  {title} {Experimental quantum teleportation},\ }\href
  {https://doi.org/10.1038/37539} {\bibfield  {journal} {\bibinfo  {journal}
  {Nature}\ }\textbf {\bibinfo {volume} {390}},\ \bibinfo {pages} {575}
  (\bibinfo {year} {1997})}\BibitemShut {NoStop}%
\bibitem [{\citenamefont {Boschi}\ \emph {et~al.}(1998)\citenamefont {Boschi},
  \citenamefont {Branca}, \citenamefont {De~Martini}, \citenamefont {Hardy},\
  and\ \citenamefont {Popescu}}]{Popescu1998teleportexperiment}%
  \BibitemOpen
  \bibfield  {author} {\bibinfo {author} {\bibfnamefont {D.}~\bibnamefont
  {Boschi}}, \bibinfo {author} {\bibfnamefont {S.}~\bibnamefont {Branca}},
  \bibinfo {author} {\bibfnamefont {F.}~\bibnamefont {De~Martini}}, \bibinfo
  {author} {\bibfnamefont {L.}~\bibnamefont {Hardy}},\ and\ \bibinfo {author}
  {\bibfnamefont {S.}~\bibnamefont {Popescu}},\ }\bibfield  {title} {\bibinfo
  {title} {Experimental realization of teleporting an unknown pure quantum
  state via dual classical and einstein-podolsky-rosen channels},\ }\href
  {https://doi.org/10.1103/PhysRevLett.80.1121} {\bibfield  {journal} {\bibinfo
   {journal} {Phys. Rev. Lett.}\ }\textbf {\bibinfo {volume} {80}},\ \bibinfo
  {pages} {1121} (\bibinfo {year} {1998})}\BibitemShut {NoStop}%
\bibitem [{\citenamefont {Alsing}\ and\ \citenamefont
  {Milburn}(2003)}]{Alsing2003teleport}%
  \BibitemOpen
  \bibfield  {author} {\bibinfo {author} {\bibfnamefont {P.~M.}\ \bibnamefont
  {Alsing}}\ and\ \bibinfo {author} {\bibfnamefont {G.~J.}\ \bibnamefont
  {Milburn}},\ }\bibfield  {title} {\bibinfo {title} {Teleportation with a
  uniformly accelerated partner},\ }\href
  {https://doi.org/10.1103/PhysRevLett.91.180404} {\bibfield  {journal}
  {\bibinfo  {journal} {Phys. Rev. Lett.}\ }\textbf {\bibinfo {volume} {91}},\
  \bibinfo {pages} {180404} (\bibinfo {year} {2003})}\BibitemShut {NoStop}%
\bibitem [{\citenamefont {Fuentes-Schuller}\ and\ \citenamefont
  {Mann}(2005)}]{Fuentes2005entanglement}%
  \BibitemOpen
  \bibfield  {author} {\bibinfo {author} {\bibfnamefont {I.}~\bibnamefont
  {Fuentes-Schuller}}\ and\ \bibinfo {author} {\bibfnamefont {R.~B.}\
  \bibnamefont {Mann}},\ }\bibfield  {title} {\bibinfo {title} {Alice falls
  into a black hole: Entanglement in noninertial frames},\ }\href
  {https://doi.org/10.1103/PhysRevLett.95.120404} {\bibfield  {journal}
  {\bibinfo  {journal} {Phys. Rev. Lett.}\ }\textbf {\bibinfo {volume} {95}},\
  \bibinfo {pages} {120404} (\bibinfo {year} {2005})}\BibitemShut {NoStop}%
\bibitem [{\citenamefont {Alsing}\ and\ \citenamefont
  {Fuentes}(2012)}]{Alsing2012review}%
  \BibitemOpen
  \bibfield  {author} {\bibinfo {author} {\bibfnamefont {P.~M.}\ \bibnamefont
  {Alsing}}\ and\ \bibinfo {author} {\bibfnamefont {I.}~\bibnamefont
  {Fuentes}},\ }\bibfield  {title} {\bibinfo {title} {Observer-dependent
  entanglement},\ }\href {http://stacks.iop.org/0264-9381/29/i=22/a=224001}
  {\bibfield  {journal} {\bibinfo  {journal} {Classical and Quantum Gravity}\
  }\textbf {\bibinfo {volume} {29}},\ \bibinfo {pages} {224001} (\bibinfo
  {year} {2012})}\BibitemShut {NoStop}%
\bibitem [{\citenamefont {Hu}\ \emph {et~al.}(2012)\citenamefont {Hu},
  \citenamefont {Lin},\ and\ \citenamefont {Louko}}]{Hu2012review}%
  \BibitemOpen
  \bibfield  {author} {\bibinfo {author} {\bibfnamefont {B.~L.}\ \bibnamefont
  {Hu}}, \bibinfo {author} {\bibfnamefont {S.-Y.}\ \bibnamefont {Lin}},\ and\
  \bibinfo {author} {\bibfnamefont {J.}~\bibnamefont {Louko}},\ }\bibfield
  {title} {\bibinfo {title} {Relativistic quantum information in
  detectors{\textendash}field interactions},\ }\href
  {https://doi.org/10.1088/0264-9381/29/22/224005} {\bibfield  {journal}
  {\bibinfo  {journal} {Classical and Quantum Gravity}\ }\textbf {\bibinfo
  {volume} {29}},\ \bibinfo {pages} {224005} (\bibinfo {year}
  {2012})}\BibitemShut {NoStop}%
\bibitem [{\citenamefont {Schützhold}\ and\ \citenamefont
  {Unruh}(2005)}]{Schutzhold2005critique}%
  \BibitemOpen
  \bibfield  {author} {\bibinfo {author} {\bibfnamefont {R.}~\bibnamefont
  {Schützhold}}\ and\ \bibinfo {author} {\bibfnamefont {W.~G.}\ \bibnamefont
  {Unruh}},\ }\href {https://doi.org/10.48550/ARXIV.QUANT-PH/0506028} {\bibinfo
  {title} {Comment on "teleportation with a uniformly accelerated partner"}}
  (\bibinfo {year} {2005})\BibitemShut {NoStop}%
\bibitem [{\citenamefont {Landulfo}\ and\ \citenamefont
  {Matsas}(2009)}]{Landulfo2009suddendeath}%
  \BibitemOpen
  \bibfield  {author} {\bibinfo {author} {\bibfnamefont {A.~G.~S.}\
  \bibnamefont {Landulfo}}\ and\ \bibinfo {author} {\bibfnamefont {G.~E.~A.}\
  \bibnamefont {Matsas}},\ }\bibfield  {title} {\bibinfo {title} {{Sudden death
  of entanglement and teleportation fidelity loss via the Unruh effect}},\
  }\href {https://doi.org/10.1103/PhysRevA.80.032315} {\bibfield  {journal}
  {\bibinfo  {journal} {Phys. Rev. A}\ }\textbf {\bibinfo {volume} {80}},\
  \bibinfo {pages} {032315} (\bibinfo {year} {2009})}\BibitemShut {NoStop}%
\bibitem [{\citenamefont {Unruh}\ and\ \citenamefont
  {Wald}(1984)}]{Unruh1984detector}%
  \BibitemOpen
  \bibfield  {author} {\bibinfo {author} {\bibfnamefont {W.~G.}\ \bibnamefont
  {Unruh}}\ and\ \bibinfo {author} {\bibfnamefont {R.~M.}\ \bibnamefont
  {Wald}},\ }\bibfield  {title} {\bibinfo {title} {What happens when an
  accelerating observer detects a rindler particle},\ }\href
  {https://doi.org/10.1103/PhysRevD.29.1047} {\bibfield  {journal} {\bibinfo
  {journal} {Phys. Rev. D}\ }\textbf {\bibinfo {volume} {29}},\ \bibinfo
  {pages} {1047} (\bibinfo {year} {1984})}\BibitemShut {NoStop}%
\bibitem [{\citenamefont {Unruh}(1976)}]{Unruh1979evaporation}%
  \BibitemOpen
  \bibfield  {author} {\bibinfo {author} {\bibfnamefont {W.~G.}\ \bibnamefont
  {Unruh}},\ }\bibfield  {title} {\bibinfo {title} {Notes on black-hole
  evaporation},\ }\href {https://doi.org/10.1103/PhysRevD.14.870} {\bibfield
  {journal} {\bibinfo  {journal} {Phys. Rev. D}\ }\textbf {\bibinfo {volume}
  {14}},\ \bibinfo {pages} {870} (\bibinfo {year} {1976})}\BibitemShut
  {NoStop}%
\bibitem [{\citenamefont {{DeWitt}}(1979)}]{DeWitt1979}%
  \BibitemOpen
  \bibfield  {author} {\bibinfo {author} {\bibfnamefont {B.~S.}\ \bibnamefont
  {{DeWitt}}},\ }\bibfield  {title} {\bibinfo {title} {{Quantum gravity: the
  new synthesis}},\ }in\ \href@noop {} {\emph {\bibinfo {booktitle} {General
  Relativity: An Einstein centenary survey}}},\ \bibinfo {editor} {edited by\
  \bibinfo {editor} {\bibfnamefont {S.~W.}\ \bibnamefont {{Hawking}}}\ and\
  \bibinfo {editor} {\bibfnamefont {W.}~\bibnamefont {{Israel}}}}\ (\bibinfo
  {year} {1979})\ pp.\ \bibinfo {pages} {680--745}\BibitemShut {NoStop}%
\bibitem [{\citenamefont {Polo-G\'omez}\ \emph {et~al.}(2022)\citenamefont
  {Polo-G\'omez}, \citenamefont {Garay},\ and\ \citenamefont
  {Mart\'{\i}n-Mart\'{\i}nez}}]{josepolo2022measurement}%
  \BibitemOpen
  \bibfield  {author} {\bibinfo {author} {\bibfnamefont {J.}~\bibnamefont
  {Polo-G\'omez}}, \bibinfo {author} {\bibfnamefont {L.~J.}\ \bibnamefont
  {Garay}},\ and\ \bibinfo {author} {\bibfnamefont {E.}~\bibnamefont
  {Mart\'{\i}n-Mart\'{\i}nez}},\ }\bibfield  {title} {\bibinfo {title} {A
  detector-based measurement theory for quantum field theory},\ }\href
  {https://doi.org/10.1103/PhysRevD.105.065003} {\bibfield  {journal} {\bibinfo
   {journal} {Phys. Rev. D}\ }\textbf {\bibinfo {volume} {105}},\ \bibinfo
  {pages} {065003} (\bibinfo {year} {2022})}\BibitemShut {NoStop}%
\bibitem [{\citenamefont {Cliche}\ and\ \citenamefont
  {Kempf}(2010)}]{Cliche2010channel}%
  \BibitemOpen
  \bibfield  {author} {\bibinfo {author} {\bibfnamefont {M.}~\bibnamefont
  {Cliche}}\ and\ \bibinfo {author} {\bibfnamefont {A.}~\bibnamefont {Kempf}},\
  }\bibfield  {title} {\bibinfo {title} {Relativistic quantum channel of
  communication through field quanta},\ }\href
  {https://doi.org/10.1103/PhysRevA.81.012330} {\bibfield  {journal} {\bibinfo
  {journal} {Phys. Rev. A}\ }\textbf {\bibinfo {volume} {81}},\ \bibinfo
  {pages} {012330} (\bibinfo {year} {2010})}\BibitemShut {NoStop}%
\bibitem [{\citenamefont {Jonsson}\ \emph {et~al.}(2018)\citenamefont
  {Jonsson}, \citenamefont {Ried}, \citenamefont {Mart{\'i}n-Mart{\'i}nez},\
  and\ \citenamefont {Kempf}}]{jonsson2018transmitting}%
  \BibitemOpen
  \bibfield  {author} {\bibinfo {author} {\bibfnamefont {R.~H.}\ \bibnamefont
  {Jonsson}}, \bibinfo {author} {\bibfnamefont {K.}~\bibnamefont {Ried}},
  \bibinfo {author} {\bibfnamefont {E.}~\bibnamefont
  {Mart{\'i}n-Mart{\'i}nez}},\ and\ \bibinfo {author} {\bibfnamefont
  {A.}~\bibnamefont {Kempf}},\ }\bibfield  {title} {\bibinfo {title}
  {Transmitting qubits through relativistic fields},\ }\href
  {https://doi.org/10.1088/1751-8121/aae78a} {\bibfield  {journal} {\bibinfo
  {journal} {Journal of Physics A: Mathematical and Theoretical}\ }\textbf
  {\bibinfo {volume} {51}},\ \bibinfo {pages} {485301} (\bibinfo {year}
  {2018})}\BibitemShut {NoStop}%
\bibitem [{\citenamefont {Simidzija}\ \emph {et~al.}(2020)\citenamefont
  {Simidzija}, \citenamefont {Ahmadzadegan}, \citenamefont {Kempf},\ and\
  \citenamefont {Mart\'{\i}n-Mart\'{\i}nez}}]{Simidzija2020transmit}%
  \BibitemOpen
  \bibfield  {author} {\bibinfo {author} {\bibfnamefont {P.}~\bibnamefont
  {Simidzija}}, \bibinfo {author} {\bibfnamefont {A.}~\bibnamefont
  {Ahmadzadegan}}, \bibinfo {author} {\bibfnamefont {A.}~\bibnamefont
  {Kempf}},\ and\ \bibinfo {author} {\bibfnamefont {E.}~\bibnamefont
  {Mart\'{\i}n-Mart\'{\i}nez}},\ }\bibfield  {title} {\bibinfo {title}
  {Transmission of quantum information through quantum fields},\ }\href
  {https://doi.org/10.1103/PhysRevD.101.036014} {\bibfield  {journal} {\bibinfo
   {journal} {Phys. Rev. D}\ }\textbf {\bibinfo {volume} {101}},\ \bibinfo
  {pages} {036014} (\bibinfo {year} {2020})}\BibitemShut {NoStop}%
\bibitem [{\citenamefont {Tjoa}\ and\ \citenamefont
  {Gallock-Yoshimura}(2022)}]{tjoa2022channel}%
  \BibitemOpen
  \bibfield  {author} {\bibinfo {author} {\bibfnamefont {E.}~\bibnamefont
  {Tjoa}}\ and\ \bibinfo {author} {\bibfnamefont {K.}~\bibnamefont
  {Gallock-Yoshimura}},\ }\bibfield  {title} {\bibinfo {title} {Channel
  capacity of relativistic quantum communication with rapid interaction},\
  }\href {https://doi.org/10.1103/PhysRevD.105.085011} {\bibfield  {journal}
  {\bibinfo  {journal} {Phys. Rev. D}\ }\textbf {\bibinfo {volume} {105}},\
  \bibinfo {pages} {085011} (\bibinfo {year} {2022})}\BibitemShut {NoStop}%
\bibitem [{\citenamefont {Gallock-Yoshimura}\ and\ \citenamefont
  {Mann}(2021)}]{Gallock-Yoshimura2021mutualinfo}%
  \BibitemOpen
  \bibfield  {author} {\bibinfo {author} {\bibfnamefont {K.}~\bibnamefont
  {Gallock-Yoshimura}}\ and\ \bibinfo {author} {\bibfnamefont {R.~B.}\
  \bibnamefont {Mann}},\ }\bibfield  {title} {\bibinfo {title} {Entangled
  detectors nonperturbatively harvest mutual information},\ }\href
  {https://doi.org/10.1103/PhysRevD.104.125017} {\bibfield  {journal} {\bibinfo
   {journal} {Phys. Rev. D}\ }\textbf {\bibinfo {volume} {104}},\ \bibinfo
  {pages} {125017} (\bibinfo {year} {2021})}\BibitemShut {NoStop}%
\bibitem [{\citenamefont {Henderson}\ \emph {et~al.}(2020)\citenamefont
  {Henderson}, \citenamefont {Belenchia}, \citenamefont {Castro-Ruiz},
  \citenamefont {Budroni}, \citenamefont {Zych}, \citenamefont {Brukner},\ and\
  \citenamefont {Mann}}]{Henderson2020temporal}%
  \BibitemOpen
  \bibfield  {author} {\bibinfo {author} {\bibfnamefont {L.~J.}\ \bibnamefont
  {Henderson}}, \bibinfo {author} {\bibfnamefont {A.}~\bibnamefont
  {Belenchia}}, \bibinfo {author} {\bibfnamefont {E.}~\bibnamefont
  {Castro-Ruiz}}, \bibinfo {author} {\bibfnamefont {C.}~\bibnamefont
  {Budroni}}, \bibinfo {author} {\bibfnamefont {M.}~\bibnamefont {Zych}},
  \bibinfo {author} {\bibfnamefont {i.~c.~v.}\ \bibnamefont {Brukner}},\ and\
  \bibinfo {author} {\bibfnamefont {R.~B.}\ \bibnamefont {Mann}},\ }\bibfield
  {title} {\bibinfo {title} {Quantum temporal superposition: The case of
  quantum field theory},\ }\href
  {https://doi.org/10.1103/PhysRevLett.125.131602} {\bibfield  {journal}
  {\bibinfo  {journal} {Phys. Rev. Lett.}\ }\textbf {\bibinfo {volume} {125}},\
  \bibinfo {pages} {131602} (\bibinfo {year} {2020})}\BibitemShut {NoStop}%
\bibitem [{\citenamefont {Simidzija}\ \emph {et~al.}(2018)\citenamefont
  {Simidzija}, \citenamefont {Jonsson},\ and\ \citenamefont
  {Mart\'{\i}n-Mart\'{\i}nez}}]{Simidzija2018nogo}%
  \BibitemOpen
  \bibfield  {author} {\bibinfo {author} {\bibfnamefont {P.}~\bibnamefont
  {Simidzija}}, \bibinfo {author} {\bibfnamefont {R.~H.}\ \bibnamefont
  {Jonsson}},\ and\ \bibinfo {author} {\bibfnamefont {E.}~\bibnamefont
  {Mart\'{\i}n-Mart\'{\i}nez}},\ }\bibfield  {title} {\bibinfo {title} {General
  no-go theorem for entanglement extraction},\ }\href
  {https://doi.org/10.1103/PhysRevD.97.125002} {\bibfield  {journal} {\bibinfo
  {journal} {Phys. Rev. D}\ }\textbf {\bibinfo {volume} {97}},\ \bibinfo
  {pages} {125002} (\bibinfo {year} {2018})}\BibitemShut {NoStop}%
\bibitem [{\citenamefont {Sahu}\ \emph {et~al.}(2022)\citenamefont {Sahu},
  \citenamefont {Melgarejo-Lermas},\ and\ \citenamefont
  {Mart\'{\i}n-Mart\'{\i}nez}}]{sahu2021sabotaging}%
  \BibitemOpen
  \bibfield  {author} {\bibinfo {author} {\bibfnamefont {A.}~\bibnamefont
  {Sahu}}, \bibinfo {author} {\bibfnamefont {I.}~\bibnamefont
  {Melgarejo-Lermas}},\ and\ \bibinfo {author} {\bibfnamefont {E.}~\bibnamefont
  {Mart\'{\i}n-Mart\'{\i}nez}},\ }\bibfield  {title} {\bibinfo {title}
  {Sabotaging the harvesting of correlations from quantum fields},\ }\href
  {https://doi.org/10.1103/PhysRevD.105.065011} {\bibfield  {journal} {\bibinfo
   {journal} {Phys. Rev. D}\ }\textbf {\bibinfo {volume} {105}},\ \bibinfo
  {pages} {065011} (\bibinfo {year} {2022})}\BibitemShut {NoStop}%
\bibitem [{\citenamefont {Vilasini}\ and\ \citenamefont
  {Renner}(2022)}]{vilasini2022embedding}%
  \BibitemOpen
  \bibfield  {author} {\bibinfo {author} {\bibfnamefont {V.}~\bibnamefont
  {Vilasini}}\ and\ \bibinfo {author} {\bibfnamefont {R.}~\bibnamefont
  {Renner}},\ }\bibfield  {title} {\bibinfo {title} {Embedding cyclic causal
  structures in acyclic spacetimes: no-go results for process matrices},\
  }\href@noop {} {\bibfield  {journal} {\bibinfo  {journal} {arXiv preprint
  arXiv:2203.11245}\ } (\bibinfo {year} {2022})}\BibitemShut {NoStop}%
\bibitem [{\citenamefont {Paunkovi{\'{c}}}\ and\ \citenamefont
  {Vojinovi{\'{c}}}(2020)}]{Paunkovic2020causalorders}%
  \BibitemOpen
  \bibfield  {author} {\bibinfo {author} {\bibfnamefont {N.}~\bibnamefont
  {Paunkovi{\'{c}}}}\ and\ \bibinfo {author} {\bibfnamefont {M.}~\bibnamefont
  {Vojinovi{\'{c}}}},\ }\bibfield  {title} {\bibinfo {title} {Causal orders,
  quantum circuits and spacetime: distinguishing between definite and
  superposed causal orders},\ }\href
  {https://doi.org/10.22331/q-2020-05-28-275} {\bibfield  {journal} {\bibinfo
  {journal} {{Quantum}}\ }\textbf {\bibinfo {volume} {4}},\ \bibinfo {pages}
  {275} (\bibinfo {year} {2020})}\BibitemShut {NoStop}%
\bibitem [{\citenamefont {Wilde}(2013)}]{Wilde2013textbook}%
  \BibitemOpen
  \bibfield  {author} {\bibinfo {author} {\bibfnamefont {M.~M.}\ \bibnamefont
  {Wilde}},\ }\href@noop {} {\emph {\bibinfo {title} {Quantum Information
  Theory}}},\ \bibinfo {edition} {1st}\ ed.\ (\bibinfo  {publisher} {Cambridge
  University Press},\ \bibinfo {address} {USA},\ \bibinfo {year}
  {2013})\BibitemShut {NoStop}%
\bibitem [{\citenamefont {Leifer}\ and\ \citenamefont
  {Spekkens}(2013)}]{Leifer2013bayesian}%
  \BibitemOpen
  \bibfield  {author} {\bibinfo {author} {\bibfnamefont {M.~S.}\ \bibnamefont
  {Leifer}}\ and\ \bibinfo {author} {\bibfnamefont {R.~W.}\ \bibnamefont
  {Spekkens}},\ }\bibfield  {title} {\bibinfo {title} {Towards a formulation of
  quantum theory as a causally neutral theory of bayesian inference},\ }\href
  {https://doi.org/10.1103/PhysRevA.88.052130} {\bibfield  {journal} {\bibinfo
  {journal} {Phys. Rev. A}\ }\textbf {\bibinfo {volume} {88}},\ \bibinfo
  {pages} {052130} (\bibinfo {year} {2013})}\BibitemShut {NoStop}%
\bibitem [{\citenamefont {Leifer}(2006)}]{Leofer2006dynamics}%
  \BibitemOpen
  \bibfield  {author} {\bibinfo {author} {\bibfnamefont {M.~S.}\ \bibnamefont
  {Leifer}},\ }\bibfield  {title} {\bibinfo {title} {Quantum dynamics as an
  analog of conditional probability},\ }\href
  {https://doi.org/10.1103/PhysRevA.74.042310} {\bibfield  {journal} {\bibinfo
  {journal} {Phys. Rev. A}\ }\textbf {\bibinfo {volume} {74}},\ \bibinfo
  {pages} {042310} (\bibinfo {year} {2006})}\BibitemShut {NoStop}%
\bibitem [{\citenamefont {Wood}\ and\ \citenamefont
  {Spekkens}(2015)}]{Wood2015causal}%
  \BibitemOpen
  \bibfield  {author} {\bibinfo {author} {\bibfnamefont {C.~J.}\ \bibnamefont
  {Wood}}\ and\ \bibinfo {author} {\bibfnamefont {R.~W.}\ \bibnamefont
  {Spekkens}},\ }\bibfield  {title} {\bibinfo {title} {The lesson of causal
  discovery algorithms for quantum correlations: causal explanations of
  bell-inequality violations require fine-tuning},\ }\href
  {https://doi.org/10.1088/1367-2630/17/3/033002} {\bibfield  {journal}
  {\bibinfo  {journal} {New Journal of Physics}\ }\textbf {\bibinfo {volume}
  {17}},\ \bibinfo {pages} {033002} (\bibinfo {year} {2015})}\BibitemShut
  {NoStop}%
\bibitem [{\citenamefont {Pienaar}\ and\ \citenamefont
  {Brukner}(2015)}]{Pienaar2015causal}%
  \BibitemOpen
  \bibfield  {author} {\bibinfo {author} {\bibfnamefont {J.}~\bibnamefont
  {Pienaar}}\ and\ \bibinfo {author} {\bibfnamefont {{\v{C}}.}~\bibnamefont
  {Brukner}},\ }\bibfield  {title} {\bibinfo {title} {A graph-separation
  theorem for quantum causal models},\ }\href
  {https://doi.org/10.1088/1367-2630/17/7/073020} {\bibfield  {journal}
  {\bibinfo  {journal} {New Journal of Physics}\ }\textbf {\bibinfo {volume}
  {17}},\ \bibinfo {pages} {073020} (\bibinfo {year} {2015})}\BibitemShut
  {NoStop}%
\bibitem [{\citenamefont {Ried}\ \emph {et~al.}(2015)\citenamefont {Ried},
  \citenamefont {Agnew}, \citenamefont {Vermeyden}, \citenamefont {Janzing},
  \citenamefont {Spekkens},\ and\ \citenamefont {Resch}}]{ried2015quantum}%
  \BibitemOpen
  \bibfield  {author} {\bibinfo {author} {\bibfnamefont {K.}~\bibnamefont
  {Ried}}, \bibinfo {author} {\bibfnamefont {M.}~\bibnamefont {Agnew}},
  \bibinfo {author} {\bibfnamefont {L.}~\bibnamefont {Vermeyden}}, \bibinfo
  {author} {\bibfnamefont {D.}~\bibnamefont {Janzing}}, \bibinfo {author}
  {\bibfnamefont {R.~W.}\ \bibnamefont {Spekkens}},\ and\ \bibinfo {author}
  {\bibfnamefont {K.~J.}\ \bibnamefont {Resch}},\ }\bibfield  {title} {\bibinfo
  {title} {A quantum advantage for inferring causal structure},\ }\href@noop {}
  {\bibfield  {journal} {\bibinfo  {journal} {Nature Physics}\ }\textbf
  {\bibinfo {volume} {11}},\ \bibinfo {pages} {414} (\bibinfo {year}
  {2015})}\BibitemShut {NoStop}%
\bibitem [{\citenamefont {Costa}\ and\ \citenamefont
  {Shrapnel}(2016)}]{Costa2016causal}%
  \BibitemOpen
  \bibfield  {author} {\bibinfo {author} {\bibfnamefont {F.}~\bibnamefont
  {Costa}}\ and\ \bibinfo {author} {\bibfnamefont {S.}~\bibnamefont
  {Shrapnel}},\ }\bibfield  {title} {\bibinfo {title} {Quantum causal
  modelling},\ }\href {https://doi.org/10.1088/1367-2630/18/6/063032}
  {\bibfield  {journal} {\bibinfo  {journal} {New Journal of Physics}\ }\textbf
  {\bibinfo {volume} {18}},\ \bibinfo {pages} {063032} (\bibinfo {year}
  {2016})}\BibitemShut {NoStop}%
\bibitem [{\citenamefont {Fritz}(2012)}]{Fritz2012beyond}%
  \BibitemOpen
  \bibfield  {author} {\bibinfo {author} {\bibfnamefont {T.}~\bibnamefont
  {Fritz}},\ }\bibfield  {title} {\bibinfo {title} {Beyond
  bell{\textquotesingle}s theorem: correlation scenarios},\ }\href
  {https://doi.org/10.1088/1367-2630/14/10/103001} {\bibfield  {journal}
  {\bibinfo  {journal} {New Journal of Physics}\ }\textbf {\bibinfo {volume}
  {14}},\ \bibinfo {pages} {103001} (\bibinfo {year} {2012})}\BibitemShut
  {NoStop}%
\bibitem [{\citenamefont {Hardy}(2007)}]{Hardy2007towards}%
  \BibitemOpen
  \bibfield  {author} {\bibinfo {author} {\bibfnamefont {L.}~\bibnamefont
  {Hardy}},\ }\bibfield  {title} {\bibinfo {title} {Towards quantum gravity: a
  framework for probabilistic theories with non-fixed causal structure},\
  }\href {https://doi.org/10.1088/1751-8113/40/12/s12} {\bibfield  {journal}
  {\bibinfo  {journal} {Journal of Physics A: Mathematical and Theoretical}\
  }\textbf {\bibinfo {volume} {40}},\ \bibinfo {pages} {3081} (\bibinfo {year}
  {2007})}\BibitemShut {NoStop}%
\bibitem [{\citenamefont {Zych}\ \emph {et~al.}(2019)\citenamefont {Zych},
  \citenamefont {Costa}, \citenamefont {Pikovski},\ and\ \citenamefont
  {Brukner}}]{zych2019bell}%
  \BibitemOpen
  \bibfield  {author} {\bibinfo {author} {\bibfnamefont {M.}~\bibnamefont
  {Zych}}, \bibinfo {author} {\bibfnamefont {F.}~\bibnamefont {Costa}},
  \bibinfo {author} {\bibfnamefont {I.}~\bibnamefont {Pikovski}},\ and\
  \bibinfo {author} {\bibfnamefont {{\v{C}}.}~\bibnamefont {Brukner}},\
  }\bibfield  {title} {\bibinfo {title} {Bell’s theorem for temporal order},\
  }\href@noop {} {\bibfield  {journal} {\bibinfo  {journal} {Nature
  communications}\ }\textbf {\bibinfo {volume} {10}},\ \bibinfo {pages} {1}
  (\bibinfo {year} {2019})}\BibitemShut {NoStop}%
\bibitem [{\citenamefont {Oreshkov}\ \emph {et~al.}(2012)\citenamefont
  {Oreshkov}, \citenamefont {Costa},\ and\ \citenamefont
  {Brukner}}]{oreshkov2012quantum}%
  \BibitemOpen
  \bibfield  {author} {\bibinfo {author} {\bibfnamefont {O.}~\bibnamefont
  {Oreshkov}}, \bibinfo {author} {\bibfnamefont {F.}~\bibnamefont {Costa}},\
  and\ \bibinfo {author} {\bibfnamefont {{\v{C}}.}~\bibnamefont {Brukner}},\
  }\bibfield  {title} {\bibinfo {title} {Quantum correlations with no causal
  order},\ }\href@noop {} {\bibfield  {journal} {\bibinfo  {journal} {Nature
  communications}\ }\textbf {\bibinfo {volume} {3}},\ \bibinfo {pages} {1}
  (\bibinfo {year} {2012})}\BibitemShut {NoStop}%
\bibitem [{\citenamefont {Chiribella}\ \emph {et~al.}(2013)\citenamefont
  {Chiribella}, \citenamefont {D'Ariano}, \citenamefont {Perinotti},\ and\
  \citenamefont {Valiron}}]{Chiribella2013computation}%
  \BibitemOpen
  \bibfield  {author} {\bibinfo {author} {\bibfnamefont {G.}~\bibnamefont
  {Chiribella}}, \bibinfo {author} {\bibfnamefont {G.~M.}\ \bibnamefont
  {D'Ariano}}, \bibinfo {author} {\bibfnamefont {P.}~\bibnamefont
  {Perinotti}},\ and\ \bibinfo {author} {\bibfnamefont {B.}~\bibnamefont
  {Valiron}},\ }\bibfield  {title} {\bibinfo {title} {Quantum computations
  without definite causal structure},\ }\href
  {https://doi.org/10.1103/PhysRevA.88.022318} {\bibfield  {journal} {\bibinfo
  {journal} {Phys. Rev. A}\ }\textbf {\bibinfo {volume} {88}},\ \bibinfo
  {pages} {022318} (\bibinfo {year} {2013})}\BibitemShut {NoStop}%
\bibitem [{\citenamefont {Ara{\'{u}}jo}\ \emph {et~al.}(2015)\citenamefont
  {Ara{\'{u}}jo}, \citenamefont {Branciard}, \citenamefont {Costa},
  \citenamefont {Feix}, \citenamefont {Giarmatzi},\ and\ \citenamefont
  {Brukner}}]{Arajo2015nonseparable}%
  \BibitemOpen
  \bibfield  {author} {\bibinfo {author} {\bibfnamefont {M.}~\bibnamefont
  {Ara{\'{u}}jo}}, \bibinfo {author} {\bibfnamefont {C.}~\bibnamefont
  {Branciard}}, \bibinfo {author} {\bibfnamefont {F.}~\bibnamefont {Costa}},
  \bibinfo {author} {\bibfnamefont {A.}~\bibnamefont {Feix}}, \bibinfo {author}
  {\bibfnamefont {C.}~\bibnamefont {Giarmatzi}},\ and\ \bibinfo {author}
  {\bibfnamefont {{\v{C}}.}~\bibnamefont {Brukner}},\ }\bibfield  {title}
  {\bibinfo {title} {Witnessing causal nonseparability},\ }\href
  {https://doi.org/10.1088/1367-2630/17/10/102001} {\bibfield  {journal}
  {\bibinfo  {journal} {New Journal of Physics}\ }\textbf {\bibinfo {volume}
  {17}},\ \bibinfo {pages} {102001} (\bibinfo {year} {2015})}\BibitemShut
  {NoStop}%
\bibitem [{\citenamefont {Oreshkov}\ and\ \citenamefont
  {Giarmatzi}(2016)}]{Oreshkov2016separable}%
  \BibitemOpen
  \bibfield  {author} {\bibinfo {author} {\bibfnamefont {O.}~\bibnamefont
  {Oreshkov}}\ and\ \bibinfo {author} {\bibfnamefont {C.}~\bibnamefont
  {Giarmatzi}},\ }\bibfield  {title} {\bibinfo {title} {Causal and causally
  separable processes},\ }\href {https://doi.org/10.1088/1367-2630/18/9/093020}
  {\bibfield  {journal} {\bibinfo  {journal} {New Journal of Physics}\ }\textbf
  {\bibinfo {volume} {18}},\ \bibinfo {pages} {093020} (\bibinfo {year}
  {2016})}\BibitemShut {NoStop}%
\bibitem [{\citenamefont {Chiribella}\ \emph {et~al.}(2021)\citenamefont
  {Chiribella}, \citenamefont {Banik}, \citenamefont {Bhattacharya},
  \citenamefont {Guha}, \citenamefont {Alimuddin}, \citenamefont {Roy},
  \citenamefont {Saha}, \citenamefont {Agrawal},\ and\ \citenamefont
  {Kar}}]{Chiribella2021comm}%
  \BibitemOpen
  \bibfield  {author} {\bibinfo {author} {\bibfnamefont {G.}~\bibnamefont
  {Chiribella}}, \bibinfo {author} {\bibfnamefont {M.}~\bibnamefont {Banik}},
  \bibinfo {author} {\bibfnamefont {S.~S.}\ \bibnamefont {Bhattacharya}},
  \bibinfo {author} {\bibfnamefont {T.}~\bibnamefont {Guha}}, \bibinfo {author}
  {\bibfnamefont {M.}~\bibnamefont {Alimuddin}}, \bibinfo {author}
  {\bibfnamefont {A.}~\bibnamefont {Roy}}, \bibinfo {author} {\bibfnamefont
  {S.}~\bibnamefont {Saha}}, \bibinfo {author} {\bibfnamefont {S.}~\bibnamefont
  {Agrawal}},\ and\ \bibinfo {author} {\bibfnamefont {G.}~\bibnamefont {Kar}},\
  }\bibfield  {title} {\bibinfo {title} {Indefinite causal order enables
  perfect quantum communication with zero capacity channels},\ }\href
  {https://doi.org/10.1088/1367-2630/abe7a0} {\bibfield  {journal} {\bibinfo
  {journal} {New Journal of Physics}\ }\textbf {\bibinfo {volume} {23}},\
  \bibinfo {pages} {033039} (\bibinfo {year} {2021})}\BibitemShut {NoStop}%
\bibitem [{\citenamefont {Felce}\ and\ \citenamefont
  {Vedral}(2020)}]{Vedral2020fridge}%
  \BibitemOpen
  \bibfield  {author} {\bibinfo {author} {\bibfnamefont {D.}~\bibnamefont
  {Felce}}\ and\ \bibinfo {author} {\bibfnamefont {V.}~\bibnamefont {Vedral}},\
  }\bibfield  {title} {\bibinfo {title} {Quantum refrigeration with indefinite
  causal order},\ }\href {https://doi.org/10.1103/PhysRevLett.125.070603}
  {\bibfield  {journal} {\bibinfo  {journal} {Phys. Rev. Lett.}\ }\textbf
  {\bibinfo {volume} {125}},\ \bibinfo {pages} {070603} (\bibinfo {year}
  {2020})}\BibitemShut {NoStop}%
\bibitem [{\citenamefont {Nielsen}\ and\ \citenamefont
  {Chuang}(2000)}]{nielsen2000quantum}%
  \BibitemOpen
  \bibfield  {author} {\bibinfo {author} {\bibfnamefont {M.}~\bibnamefont
  {Nielsen}}\ and\ \bibinfo {author} {\bibfnamefont {I.}~\bibnamefont
  {Chuang}},\ }\href {https://books.google.ca/books?id=65FqEKQOfP8C} {\emph
  {\bibinfo {title} {Quantum Computation and Quantum Information}}},\ Cambridge
  Series on Information and the Natural Sciences\ (\bibinfo  {publisher}
  {Cambridge University Press},\ \bibinfo {year} {2000})\BibitemShut {NoStop}%
\bibitem [{\citenamefont {Holevo}(1998)}]{holevo1998capacity}%
  \BibitemOpen
  \bibfield  {author} {\bibinfo {author} {\bibfnamefont {A.~S.}\ \bibnamefont
  {Holevo}},\ }\bibfield  {title} {\bibinfo {title} {The capacity of the
  quantum channel with general signal states},\ }\href@noop {} {\bibfield
  {journal} {\bibinfo  {journal} {IEEE Transactions on Information Theory}\
  }\textbf {\bibinfo {volume} {44}},\ \bibinfo {pages} {269} (\bibinfo {year}
  {1998})}\BibitemShut {NoStop}%
\bibitem [{\citenamefont {Schumacher}\ and\ \citenamefont
  {Westmoreland}(1997)}]{schumacher1997sending}%
  \BibitemOpen
  \bibfield  {author} {\bibinfo {author} {\bibfnamefont {B.}~\bibnamefont
  {Schumacher}}\ and\ \bibinfo {author} {\bibfnamefont {M.~D.}\ \bibnamefont
  {Westmoreland}},\ }\bibfield  {title} {\bibinfo {title} {Sending classical
  information via noisy quantum channels},\ }\href@noop {} {\bibfield
  {journal} {\bibinfo  {journal} {Physical Review A}\ }\textbf {\bibinfo
  {volume} {56}},\ \bibinfo {pages} {131} (\bibinfo {year} {1997})}\BibitemShut
  {NoStop}%
\bibitem [{\citenamefont {Hayashi}\ and\ \citenamefont
  {Nagaoka}(2003)}]{hayashi2003general}%
  \BibitemOpen
  \bibfield  {author} {\bibinfo {author} {\bibfnamefont {M.}~\bibnamefont
  {Hayashi}}\ and\ \bibinfo {author} {\bibfnamefont {H.}~\bibnamefont
  {Nagaoka}},\ }\bibfield  {title} {\bibinfo {title} {General formulas for
  capacity of classical-quantum channels},\ }\href@noop {} {\bibfield
  {journal} {\bibinfo  {journal} {IEEE Transactions on Information Theory}\
  }\textbf {\bibinfo {volume} {49}},\ \bibinfo {pages} {1753} (\bibinfo {year}
  {2003})}\BibitemShut {NoStop}%
\bibitem [{\citenamefont {Kretschmann}\ and\ \citenamefont
  {Werner}(2005)}]{kretschmann2005quantum}%
  \BibitemOpen
  \bibfield  {author} {\bibinfo {author} {\bibfnamefont {D.}~\bibnamefont
  {Kretschmann}}\ and\ \bibinfo {author} {\bibfnamefont {R.~F.}\ \bibnamefont
  {Werner}},\ }\bibfield  {title} {\bibinfo {title} {Quantum channels with
  memory},\ }\href@noop {} {\bibfield  {journal} {\bibinfo  {journal} {Physical
  Review A}\ }\textbf {\bibinfo {volume} {72}},\ \bibinfo {pages} {062323}
  (\bibinfo {year} {2005})}\BibitemShut {NoStop}%
\bibitem [{\citenamefont {Wang}\ and\ \citenamefont
  {Renner}(2012)}]{Renner2012oneshot}%
  \BibitemOpen
  \bibfield  {author} {\bibinfo {author} {\bibfnamefont {L.}~\bibnamefont
  {Wang}}\ and\ \bibinfo {author} {\bibfnamefont {R.}~\bibnamefont {Renner}},\
  }\bibfield  {title} {\bibinfo {title} {One-shot classical-quantum capacity
  and hypothesis testing},\ }\href
  {https://doi.org/10.1103/PhysRevLett.108.200501} {\bibfield  {journal}
  {\bibinfo  {journal} {Phys. Rev. Lett.}\ }\textbf {\bibinfo {volume} {108}},\
  \bibinfo {pages} {200501} (\bibinfo {year} {2012})}\BibitemShut {NoStop}%
\bibitem [{\citenamefont {Buscemi}\ and\ \citenamefont
  {Datta}(2010)}]{Buscemi2010oneshot}%
  \BibitemOpen
  \bibfield  {author} {\bibinfo {author} {\bibfnamefont {F.}~\bibnamefont
  {Buscemi}}\ and\ \bibinfo {author} {\bibfnamefont {N.}~\bibnamefont
  {Datta}},\ }\bibfield  {title} {\bibinfo {title} {The quantum capacity of
  channels with arbitrarily correlated noise},\ }\href
  {https://doi.org/10.1109/TIT.2009.2039166} {\bibfield  {journal} {\bibinfo
  {journal} {IEEE Transactions on Information Theory}\ }\textbf {\bibinfo
  {volume} {56}},\ \bibinfo {pages} {1447} (\bibinfo {year}
  {2010})}\BibitemShut {NoStop}%
\bibitem [{\citenamefont {Datta}\ and\ \citenamefont
  {Hsieh}(2013)}]{Datta2013oneshotassist}%
  \BibitemOpen
  \bibfield  {author} {\bibinfo {author} {\bibfnamefont {N.}~\bibnamefont
  {Datta}}\ and\ \bibinfo {author} {\bibfnamefont {M.-H.}\ \bibnamefont
  {Hsieh}},\ }\bibfield  {title} {\bibinfo {title} {One-shot
  entanglement-assisted quantum and classical communication},\ }\href
  {https://doi.org/10.1109/TIT.2012.2228737} {\bibfield  {journal} {\bibinfo
  {journal} {IEEE Transactions on Information Theory}\ }\textbf {\bibinfo
  {volume} {59}},\ \bibinfo {pages} {1929} (\bibinfo {year}
  {2013})}\BibitemShut {NoStop}%
\bibitem [{\citenamefont {Datta}\ \emph {et~al.}(2013)\citenamefont {Datta},
  \citenamefont {Mosonyi}, \citenamefont {Hsieh},\ and\ \citenamefont
  {Brandão}}]{Datta2013smooth}%
  \BibitemOpen
  \bibfield  {author} {\bibinfo {author} {\bibfnamefont {N.}~\bibnamefont
  {Datta}}, \bibinfo {author} {\bibfnamefont {M.}~\bibnamefont {Mosonyi}},
  \bibinfo {author} {\bibfnamefont {M.-H.}\ \bibnamefont {Hsieh}},\ and\
  \bibinfo {author} {\bibfnamefont {F.~G. S.~L.}\ \bibnamefont {Brandão}},\
  }\bibfield  {title} {\bibinfo {title} {A smooth entropy approach to quantum
  hypothesis testing and the classical capacity of quantum channels},\ }\href
  {https://doi.org/10.1109/TIT.2013.2282160} {\bibfield  {journal} {\bibinfo
  {journal} {IEEE Transactions on Information Theory}\ }\textbf {\bibinfo
  {volume} {59}},\ \bibinfo {pages} {8014} (\bibinfo {year}
  {2013})}\BibitemShut {NoStop}%
\bibitem [{\citenamefont {Anshu}\ \emph {et~al.}(2019)\citenamefont {Anshu},
  \citenamefont {Jain},\ and\ \citenamefont {Warsi}}]{Anshu2019oneshot}%
  \BibitemOpen
  \bibfield  {author} {\bibinfo {author} {\bibfnamefont {A.}~\bibnamefont
  {Anshu}}, \bibinfo {author} {\bibfnamefont {R.}~\bibnamefont {Jain}},\ and\
  \bibinfo {author} {\bibfnamefont {N.~A.}\ \bibnamefont {Warsi}},\ }\bibfield
  {title} {\bibinfo {title} {On the near-optimality of one-shot classical
  communication over quantum channels},\ }\href
  {https://doi.org/10.1063/1.5039796} {\bibfield  {journal} {\bibinfo
  {journal} {Journal of Mathematical Physics}\ }\textbf {\bibinfo {volume}
  {60}},\ \bibinfo {pages} {012204} (\bibinfo {year} {2019})},\ \Eprint
  {https://arxiv.org/abs/https://doi.org/10.1063/1.5039796}
  {https://doi.org/10.1063/1.5039796} \BibitemShut {NoStop}%
\bibitem [{\citenamefont {Hiai}\ and\ \citenamefont
  {Petz}(1991)}]{hiai1991proper}%
  \BibitemOpen
  \bibfield  {author} {\bibinfo {author} {\bibfnamefont {F.}~\bibnamefont
  {Hiai}}\ and\ \bibinfo {author} {\bibfnamefont {D.}~\bibnamefont {Petz}},\
  }\bibfield  {title} {\bibinfo {title} {The proper formula for relative
  entropy and its asymptotics in quantum probability},\ }\href@noop {}
  {\bibfield  {journal} {\bibinfo  {journal} {Communications in mathematical
  physics}\ }\textbf {\bibinfo {volume} {143}},\ \bibinfo {pages} {99}
  (\bibinfo {year} {1991})}\BibitemShut {NoStop}%
\bibitem [{\citenamefont {Landulfo}(2016)}]{Landulfo2016magnus1}%
  \BibitemOpen
  \bibfield  {author} {\bibinfo {author} {\bibfnamefont {A.~G.~S.}\
  \bibnamefont {Landulfo}},\ }\bibfield  {title} {\bibinfo {title}
  {Nonperturbative approach to relativistic quantum communication channels},\
  }\href {https://doi.org/10.1103/PhysRevD.93.104019} {\bibfield  {journal}
  {\bibinfo  {journal} {Phys. Rev. D}\ }\textbf {\bibinfo {volume} {93}},\
  \bibinfo {pages} {104019} (\bibinfo {year} {2016})}\BibitemShut {NoStop}%
\bibitem [{\citenamefont {Barcellos}\ and\ \citenamefont
  {Landulfo}(2021)}]{Landulfo2021cost}%
  \BibitemOpen
  \bibfield  {author} {\bibinfo {author} {\bibfnamefont {I.~B.}\ \bibnamefont
  {Barcellos}}\ and\ \bibinfo {author} {\bibfnamefont {A.~G.~S.}\ \bibnamefont
  {Landulfo}},\ }\bibfield  {title} {\bibinfo {title} {Relativistic quantum
  communication: Energy cost and channel capacities},\ }\href
  {https://doi.org/10.1103/PhysRevD.104.105018} {\bibfield  {journal} {\bibinfo
   {journal} {Phys. Rev. D}\ }\textbf {\bibinfo {volume} {104}},\ \bibinfo
  {pages} {105018} (\bibinfo {year} {2021})}\BibitemShut {NoStop}%
\bibitem [{\citenamefont {Mart\'{\i}n-Mart\'{\i}nez}\ \emph
  {et~al.}(2020)\citenamefont {Mart\'{\i}n-Mart\'{\i}nez}, \citenamefont
  {Perche},\ and\ \citenamefont {de~S.~L.~Torres}}]{Tales2020GRQO}%
  \BibitemOpen
  \bibfield  {author} {\bibinfo {author} {\bibfnamefont {E.}~\bibnamefont
  {Mart\'{\i}n-Mart\'{\i}nez}}, \bibinfo {author} {\bibfnamefont {T.~R.}\
  \bibnamefont {Perche}},\ and\ \bibinfo {author} {\bibfnamefont
  {B.}~\bibnamefont {de~S.~L.~Torres}},\ }\bibfield  {title} {\bibinfo {title}
  {General relativistic quantum optics: Finite-size particle detector models in
  curved spacetimes},\ }\href {https://doi.org/10.1103/PhysRevD.101.045017}
  {\bibfield  {journal} {\bibinfo  {journal} {Phys. Rev. D}\ }\textbf {\bibinfo
  {volume} {101}},\ \bibinfo {pages} {045017} (\bibinfo {year}
  {2020})}\BibitemShut {NoStop}%
\bibitem [{\citenamefont {Mart\'{\i}n-Mart\'{\i}nez}\ \emph
  {et~al.}(2021)\citenamefont {Mart\'{\i}n-Mart\'{\i}nez}, \citenamefont
  {Perche},\ and\ \citenamefont {de~S.~L.~Torres}}]{Bruno2020time-ordering}%
  \BibitemOpen
  \bibfield  {author} {\bibinfo {author} {\bibfnamefont {E.}~\bibnamefont
  {Mart\'{\i}n-Mart\'{\i}nez}}, \bibinfo {author} {\bibfnamefont {T.~R.}\
  \bibnamefont {Perche}},\ and\ \bibinfo {author} {\bibfnamefont
  {B.}~\bibnamefont {de~S.~L.~Torres}},\ }\bibfield  {title} {\bibinfo {title}
  {Broken covariance of particle detector models in relativistic quantum
  information},\ }\href {https://doi.org/10.1103/PhysRevD.103.025007}
  {\bibfield  {journal} {\bibinfo  {journal} {Phys. Rev. D}\ }\textbf {\bibinfo
  {volume} {103}},\ \bibinfo {pages} {025007} (\bibinfo {year}
  {2021})}\BibitemShut {NoStop}%
\bibitem [{\citenamefont {Perche}(2022)}]{Tales2022FNC}%
  \BibitemOpen
  \bibfield  {author} {\bibinfo {author} {\bibfnamefont {T.~R.}\ \bibnamefont
  {Perche}},\ }\bibfield  {title} {\bibinfo {title} {Localized non-relativistic
  quantum systems in curved spacetimes: a general characterization of particle
  detector models},\ }\href {https://arxiv.org/abs/2206.01225} {\bibfield
  {journal} {\bibinfo  {journal} {arxiv:2206.01225}\ } (\bibinfo {year}
  {2022})}\BibitemShut {NoStop}%
\bibitem [{\citenamefont {Tjoa}(2022)}]{Tjoa2022fermi}%
  \BibitemOpen
  \bibfield  {author} {\bibinfo {author} {\bibfnamefont {E.}~\bibnamefont
  {Tjoa}},\ }\bibfield  {title} {\bibinfo {title} {{Fermi two-atom problem:
  non-perturbative approach via relativistic quantum information and algebraic
  quantum field theory}},\ }\href {https://arxiv.org/abs/2206.02316} {\bibfield
   {journal} {\bibinfo  {journal} {arxiv:2206.02316}\ } (\bibinfo {year}
  {2022})}\BibitemShut {NoStop}%
\bibitem [{\citenamefont {Hollands}\ and\ \citenamefont
  {Sanders}(2017)}]{hollands2017entanglement}%
  \BibitemOpen
  \bibfield  {author} {\bibinfo {author} {\bibfnamefont {S.}~\bibnamefont
  {Hollands}}\ and\ \bibinfo {author} {\bibfnamefont {K.}~\bibnamefont
  {Sanders}},\ }\bibfield  {title} {\bibinfo {title} {Entanglement measures and
  their properties in quantum field theory},\ }\bibfield  {journal} {\bibinfo
  {journal} {arXiv:1702.04924}\ }\href
  {https://doi.org/https://arxiv.org/abs/1702.04924}
  {https://arxiv.org/abs/1702.04924} (\bibinfo {year} {2017})\BibitemShut
  {NoStop}%
\bibitem [{\citenamefont {Perche}\ and\ \citenamefont
  {Shalabi}(2022)}]{perche2022spacetime}%
  \BibitemOpen
  \bibfield  {author} {\bibinfo {author} {\bibfnamefont {T.~R.}\ \bibnamefont
  {Perche}}\ and\ \bibinfo {author} {\bibfnamefont {A.}~\bibnamefont
  {Shalabi}},\ }\href@noop {} {\bibinfo {title} {Spacetime curvature from ultra
  rapid measurements of quantum fields}} (\bibinfo {year} {2022}),\ \Eprint
  {https://arxiv.org/abs/2202.11108} {arXiv:2202.11108 [quant-ph]} \BibitemShut
  {NoStop}%
\bibitem [{\citenamefont {Dimock}(1980)}]{dimock1980algebras}%
  \BibitemOpen
  \bibfield  {author} {\bibinfo {author} {\bibfnamefont {J.}~\bibnamefont
  {Dimock}},\ }\bibfield  {title} {\bibinfo {title} {Algebras of local
  observables on a manifold},\ }\href {https://doi.org/10.1007/BF01269921}
  {\bibfield  {journal} {\bibinfo  {journal} {Communications in Mathematical
  Physics}\ }\textbf {\bibinfo {volume} {77}},\ \bibinfo {pages} {219}
  (\bibinfo {year} {1980})}\BibitemShut {NoStop}%
\bibitem [{\citenamefont {Kay}\ and\ \citenamefont
  {Wald}(1991)}]{KayWald1991theorems}%
  \BibitemOpen
  \bibfield  {author} {\bibinfo {author} {\bibfnamefont {B.~S.}\ \bibnamefont
  {Kay}}\ and\ \bibinfo {author} {\bibfnamefont {R.~M.}\ \bibnamefont {Wald}},\
  }\bibfield  {title} {\bibinfo {title} {Theorems on the uniqueness and thermal
  properties of stationary, nonsingular, quasifree states on spacetimes with a
  bifurcate killing horizon},\ }\href
  {https://doi.org/https://doi.org/10.1016/0370-1573(91)90015-E} {\bibfield
  {journal} {\bibinfo  {journal} {Physics Reports}\ }\textbf {\bibinfo {volume}
  {207}},\ \bibinfo {pages} {49} (\bibinfo {year} {1991})}\BibitemShut
  {NoStop}%
\bibitem [{\citenamefont {Verdon-Akzam}\ \emph {et~al.}(2016)\citenamefont
  {Verdon-Akzam}, \citenamefont {Mart\'{\i}n-Mart\'{\i}nez},\ and\
  \citenamefont {Kempf}}]{qet2016qudit}%
  \BibitemOpen
  \bibfield  {author} {\bibinfo {author} {\bibfnamefont {G.}~\bibnamefont
  {Verdon-Akzam}}, \bibinfo {author} {\bibfnamefont {E.}~\bibnamefont
  {Mart\'{\i}n-Mart\'{\i}nez}},\ and\ \bibinfo {author} {\bibfnamefont
  {A.}~\bibnamefont {Kempf}},\ }\bibfield  {title} {\bibinfo {title}
  {Asymptotically limitless quantum energy teleportation via qudit probes},\
  }\href {https://doi.org/10.1103/PhysRevA.93.022308} {\bibfield  {journal}
  {\bibinfo  {journal} {Phys. Rev. A}\ }\textbf {\bibinfo {volume} {93}},\
  \bibinfo {pages} {022308} (\bibinfo {year} {2016})}\BibitemShut {NoStop}%
\bibitem [{\citenamefont {Gottesman}\ and\ \citenamefont
  {Chuang}(1999)}]{gottesman1999demonstrating}%
  \BibitemOpen
  \bibfield  {author} {\bibinfo {author} {\bibfnamefont {D.}~\bibnamefont
  {Gottesman}}\ and\ \bibinfo {author} {\bibfnamefont {I.~L.}\ \bibnamefont
  {Chuang}},\ }\bibfield  {title} {\bibinfo {title} {Demonstrating the
  viability of universal quantum computation using teleportation and
  single-qubit operations},\ }\href@noop {} {\bibfield  {journal} {\bibinfo
  {journal} {Nature}\ }\textbf {\bibinfo {volume} {402}},\ \bibinfo {pages}
  {390} (\bibinfo {year} {1999})}\BibitemShut {NoStop}%
\bibitem [{\citenamefont {Jozsa}(2006)}]{jozsa2006introduction}%
  \BibitemOpen
  \bibfield  {author} {\bibinfo {author} {\bibfnamefont {R.}~\bibnamefont
  {Jozsa}},\ }\bibfield  {title} {\bibinfo {title} {An introduction to
  measurement based quantum computation},\ }\href@noop {} {\bibfield  {journal}
  {\bibinfo  {journal} {NATO Science Series, III: Computer and Systems
  Sciences. Quantum Information Processing-From Theory to Experiment}\ }\textbf
  {\bibinfo {volume} {199}},\ \bibinfo {pages} {137} (\bibinfo {year}
  {2006})}\BibitemShut {NoStop}%
\bibitem [{\citenamefont {Bennett}\ \emph {et~al.}(2001)\citenamefont
  {Bennett}, \citenamefont {DiVincenzo}, \citenamefont {Shor}, \citenamefont
  {Smolin}, \citenamefont {Terhal},\ and\ \citenamefont
  {Wootters}}]{Bennett2001remote}%
  \BibitemOpen
  \bibfield  {author} {\bibinfo {author} {\bibfnamefont {C.~H.}\ \bibnamefont
  {Bennett}}, \bibinfo {author} {\bibfnamefont {D.~P.}\ \bibnamefont
  {DiVincenzo}}, \bibinfo {author} {\bibfnamefont {P.~W.}\ \bibnamefont
  {Shor}}, \bibinfo {author} {\bibfnamefont {J.~A.}\ \bibnamefont {Smolin}},
  \bibinfo {author} {\bibfnamefont {B.~M.}\ \bibnamefont {Terhal}},\ and\
  \bibinfo {author} {\bibfnamefont {W.~K.}\ \bibnamefont {Wootters}},\
  }\bibfield  {title} {\bibinfo {title} {Remote state preparation},\ }\href
  {https://doi.org/10.1103/PhysRevLett.87.077902} {\bibfield  {journal}
  {\bibinfo  {journal} {Phys. Rev. Lett.}\ }\textbf {\bibinfo {volume} {87}},\
  \bibinfo {pages} {077902} (\bibinfo {year} {2001})}\BibitemShut {NoStop}%
\bibitem [{\citenamefont {Bennett}\ \emph {et~al.}(2002)\citenamefont
  {Bennett}, \citenamefont {DiVincenzo}, \citenamefont {Shor}, \citenamefont
  {Smolin}, \citenamefont {Terhal},\ and\ \citenamefont
  {Wootters}}]{Bennett2002erratum}%
  \BibitemOpen
  \bibfield  {author} {\bibinfo {author} {\bibfnamefont {C.~H.}\ \bibnamefont
  {Bennett}}, \bibinfo {author} {\bibfnamefont {D.~P.}\ \bibnamefont
  {DiVincenzo}}, \bibinfo {author} {\bibfnamefont {P.~W.}\ \bibnamefont
  {Shor}}, \bibinfo {author} {\bibfnamefont {J.~A.}\ \bibnamefont {Smolin}},
  \bibinfo {author} {\bibfnamefont {B.~M.}\ \bibnamefont {Terhal}},\ and\
  \bibinfo {author} {\bibfnamefont {W.~K.}\ \bibnamefont {Wootters}},\
  }\bibfield  {title} {\bibinfo {title} {Erratum: Remote state preparation
  [phys. rev. lett. 87, 077902 (2001)]},\ }\href
  {https://doi.org/10.1103/PhysRevLett.88.099902} {\bibfield  {journal}
  {\bibinfo  {journal} {Phys. Rev. Lett.}\ }\textbf {\bibinfo {volume} {88}},\
  \bibinfo {pages} {099902} (\bibinfo {year} {2002})}\BibitemShut {NoStop}%
\bibitem [{\citenamefont {Fewster}\ and\ \citenamefont
  {Verch}(2020)}]{fewster2020quantum}%
  \BibitemOpen
  \bibfield  {author} {\bibinfo {author} {\bibfnamefont {C.~J.}\ \bibnamefont
  {Fewster}}\ and\ \bibinfo {author} {\bibfnamefont {R.}~\bibnamefont
  {Verch}},\ }\bibfield  {title} {\bibinfo {title} {Quantum fields and local
  measurements},\ }\href@noop {} {\bibfield  {journal} {\bibinfo  {journal}
  {Communications in Mathematical physics}\ }\textbf {\bibinfo {volume}
  {378}},\ \bibinfo {pages} {851} (\bibinfo {year} {2020})}\BibitemShut
  {NoStop}%
\bibitem [{\citenamefont {VERCH}\ and\ \citenamefont
  {WERNER}(2005)}]{Verch2005distill}%
  \BibitemOpen
  \bibfield  {author} {\bibinfo {author} {\bibfnamefont {R.}~\bibnamefont
  {VERCH}}\ and\ \bibinfo {author} {\bibfnamefont {R.~F.}\ \bibnamefont
  {WERNER}},\ }\bibfield  {title} {\bibinfo {title} {Distillability and
  positivity of partial transposes in general quantum field systems},\ }\href
  {https://doi.org/10.1142/S0129055X05002364} {\bibfield  {journal} {\bibinfo
  {journal} {Reviews in Mathematical Physics}\ }\textbf {\bibinfo {volume}
  {17}},\ \bibinfo {pages} {545} (\bibinfo {year} {2005})},\ \Eprint
  {https://arxiv.org/abs/https://doi.org/10.1142/S0129055X05002364}
  {https://doi.org/10.1142/S0129055X05002364} \BibitemShut {NoStop}%
\bibitem [{\citenamefont {Fewster}\ and\ \citenamefont
  {Rejzner}(2019)}]{fewster2020algebraic}%
  \BibitemOpen
  \bibfield  {author} {\bibinfo {author} {\bibfnamefont {C.~J.}\ \bibnamefont
  {Fewster}}\ and\ \bibinfo {author} {\bibfnamefont {K.}~\bibnamefont
  {Rejzner}},\ }\href@noop {} {\bibinfo {title} {Algebraic quantum field theory
  -- an introduction}} (\bibinfo {year} {2019}),\ \Eprint
  {https://arxiv.org/abs/1904.04051} {arXiv:1904.04051 [hep-th]} \BibitemShut
  {NoStop}%
\bibitem [{\citenamefont {Khavkine}\ and\ \citenamefont
  {Moretti}(2015)}]{Khavkhine2015AQFT}%
  \BibitemOpen
  \bibfield  {author} {\bibinfo {author} {\bibfnamefont {I.}~\bibnamefont
  {Khavkine}}\ and\ \bibinfo {author} {\bibfnamefont {V.}~\bibnamefont
  {Moretti}},\ }\bibfield  {title} {\bibinfo {title} {Algebraic {QFT} in curved
  spacetime and quasifree {Hadamard} states: An introduction},\ }\href
  {https://doi.org/10.1007/978-3-319-21353-8_5} {\bibfield  {journal} {\bibinfo
   {journal} {Mathematical Physics Studies}\ ,\ \bibinfo {pages} {191–251}}
  (\bibinfo {year} {2015})}\BibitemShut {NoStop}%
\bibitem [{\citenamefont {Radzikowski}(1996)}]{Radzikowski1996microlocal}%
  \BibitemOpen
  \bibfield  {author} {\bibinfo {author} {\bibfnamefont {M.~J.}\ \bibnamefont
  {Radzikowski}},\ }\bibfield  {title} {\bibinfo {title} {{Micro-local approach
  to the Hadamard condition in quantum field theory on curved space-time}},\
  }\href {https://doi.org/cmp/1104287114} {\bibfield  {journal} {\bibinfo
  {journal} {Communications in Mathematical Physics}\ }\textbf {\bibinfo
  {volume} {179}},\ \bibinfo {pages} {529 } (\bibinfo {year}
  {1996})}\BibitemShut {NoStop}%
\bibitem [{\citenamefont {Poisson}(2009)}]{Poisson:2009pwt}%
  \BibitemOpen
  \bibfield  {author} {\bibinfo {author} {\bibfnamefont {E.}~\bibnamefont
  {Poisson}},\ }\href {https://doi.org/10.1017/CBO9780511606601} {\emph
  {\bibinfo {title} {{A Relativist's Toolkit: The Mathematics of Black-Hole
  Mechanics}}}}\ (\bibinfo  {publisher} {Cambridge University Press},\ \bibinfo
  {year} {2009})\BibitemShut {NoStop}%
\bibitem [{\citenamefont {Wald}(2010)}]{wald2010general}%
  \BibitemOpen
  \bibfield  {author} {\bibinfo {author} {\bibfnamefont {R.}~\bibnamefont
  {Wald}},\ }\href {https://books.google.ca/books?id=9S-hzg6-moYC} {\emph
  {\bibinfo {title} {General Relativity}}}\ (\bibinfo  {publisher} {University
  of Chicago Press},\ \bibinfo {year} {2010})\BibitemShut {NoStop}%
\bibitem [{\citenamefont {Wald}\ and\ \citenamefont
  {Pfister}(1994)}]{wald1994quantum}%
  \BibitemOpen
  \bibfield  {author} {\bibinfo {author} {\bibfnamefont {R.}~\bibnamefont
  {Wald}}\ and\ \bibinfo {author} {\bibfnamefont {J.}~\bibnamefont {Pfister}},\
  }\href {https://books.google.ca/books?id=Iud7eyDxT1AC} {\emph {\bibinfo
  {title} {Quantum Field Theory in Curved Spacetime and Black Hole
  Thermodynamics}}},\ Chicago Lectures in Physics\ (\bibinfo  {publisher}
  {University of Chicago Press},\ \bibinfo {year} {1994})\BibitemShut {NoStop}%
\bibitem [{\citenamefont {Birrell}\ \emph {et~al.}(1984)\citenamefont
  {Birrell}, \citenamefont {Birrell},\ and\ \citenamefont
  {Davies}}]{birrell1984quantum}%
  \BibitemOpen
  \bibfield  {author} {\bibinfo {author} {\bibfnamefont {N.}~\bibnamefont
  {Birrell}}, \bibinfo {author} {\bibfnamefont {N.}~\bibnamefont {Birrell}},\
  and\ \bibinfo {author} {\bibfnamefont {P.}~\bibnamefont {Davies}},\ }\href
  {https://books.google.ca/books?id=SEnaUnrqzrUC} {\emph {\bibinfo {title}
  {Quantum Fields in Curved Space}}},\ Cambridge Monographs on Mathematical
  Physics\ (\bibinfo  {publisher} {Cambridge University Press},\ \bibinfo
  {year} {1984})\BibitemShut {NoStop}%
\bibitem [{\citenamefont {DeWitt}\ and\ \citenamefont
  {Brehme}(1960)}]{DeWitt1960radiation}%
  \BibitemOpen
  \bibfield  {author} {\bibinfo {author} {\bibfnamefont {B.~S.}\ \bibnamefont
  {DeWitt}}\ and\ \bibinfo {author} {\bibfnamefont {R.~W.}\ \bibnamefont
  {Brehme}},\ }\bibfield  {title} {\bibinfo {title} {Radiation damping in a
  gravitational field},\ }\href
  {https://doi.org/https://doi.org/10.1016/0003-4916(60)90030-0} {\bibfield
  {journal} {\bibinfo  {journal} {Annals of Physics}\ }\textbf {\bibinfo
  {volume} {9}},\ \bibinfo {pages} {220} (\bibinfo {year} {1960})}\BibitemShut
  {NoStop}%
\bibitem [{\citenamefont {Simidzija}\ and\ \citenamefont
  {Mart\'{\i}n-Mart\'{\i}nez}(2018)}]{simidzija2018harvesting}%
  \BibitemOpen
  \bibfield  {author} {\bibinfo {author} {\bibfnamefont {P.}~\bibnamefont
  {Simidzija}}\ and\ \bibinfo {author} {\bibfnamefont {E.}~\bibnamefont
  {Mart\'{\i}n-Mart\'{\i}nez}},\ }\bibfield  {title} {\bibinfo {title}
  {Harvesting correlations from thermal and squeezed coherent states},\ }\href
  {https://doi.org/10.1103/PhysRevD.98.085007} {\bibfield  {journal} {\bibinfo
  {journal} {Phys. Rev. D}\ }\textbf {\bibinfo {volume} {98}},\ \bibinfo
  {pages} {085007} (\bibinfo {year} {2018})}\BibitemShut {NoStop}%
\bibitem [{\citenamefont {Weldon}(2000)}]{Weldon2000thermal}%
  \BibitemOpen
  \bibfield  {author} {\bibinfo {author} {\bibfnamefont {H.~A.}\ \bibnamefont
  {Weldon}},\ }\bibfield  {title} {\bibinfo {title} {{Thermal Green functions
  in coordinate space for massless particles of any spin}},\ }\href
  {https://doi.org/10.1103/PhysRevD.62.056010} {\bibfield  {journal} {\bibinfo
  {journal} {Phys. Rev. D}\ }\textbf {\bibinfo {volume} {62}},\ \bibinfo
  {pages} {056010} (\bibinfo {year} {2000})}\BibitemShut {NoStop}%
\end{thebibliography}%

\end{document}